 \documentclass[10pt,preprint]{aastex}  %compact preprint style (MJA)
 \usepackage{natbib}
\def\ang{\AA}
\def\arcsec{\hbox{$^{\prime\prime}$}}

\def\gapprox{\lower.4ex\hbox{$\;\buildrel >\over{\scriptstyle\sim}\;$}}
\def\lapprox{\lower.4ex\hbox{$\;\buildrel <\over{\scriptstyle\sim}\;$}}

\shortauthors{ASCHWANDEN 2012}
\shorttitle{Spatio-Temporal Flare Evolution. I.}

\begin{document}
%{\sl  Manuscript, accepted ... }

\title{         Multi-Wavelength Observations of the Spatio-Temporal 
		Evolution of Solar Flares with AIA/SDO: 
		I. Universal Scaling Laws of Space and Time Parameters}

\author{        Markus J. Aschwanden }

\affil{		Lockheed Martin Advanced Technology Center,
                Org. ADBS, Bldg.252,
                3251 Hanover St.,
                Palo Alto, CA 94304, USA;
                e-mail: aschwanden@lmsal.com }
\and

\author{	Jie Zhang and Kai Liu}

\affil{ 	School of Physics, Astronomy and Computational Sciences,
		George Mason University,
		4400 University Dr., MSN 6A2,
		Fairfax, VA 22030, USA;
		e-mail: jzhang7@gmu.edu } 

\begin{abstract}
We extend a previous statistical solar flare study of 155 GOES 
M- and X-class flares observed with AIA/SDO (Aschwanden 2012)
to all 7 coronal wavelengths (94, 131, 171, 193, 211, 304, 335 \ang )
to test the wavelength-dependence of scaling laws and statistical 
distributions. Except for the 171 and 193 \ang\ wavelengths, which
are affected by EUV dimming caused by coronal mass ejections (CMEs), 
we find near-identical
size distributions of geometric (lengths $L$, flare areas $A$, volumes
$V$, fractal dimension $D_2$), temporal (flare durations $T$), and
spatio-temporal parameters (diffusion coefficient $\kappa$, spreading
exponent $\beta$, and maximum expansion velocities $v_{max}$) in
different wavelengths, which are consistent with the universal
predictions of the fractal-diffusive avalanche model of a 
slowly-driven self-organized criticality (FD-SOC) system, 
i.e., $N(L) \propto L^{-3}$, $N(A) \propto A^{-2}$, 
$N(V) \propto V^{-5/3}$, $N(T) \propto T^{-2}$,
$D_2=3/2$, for a Euclidean dimension $d=3$. 
Empirically we find also a new strong correlation $\kappa \propto 
L^{0.94\pm0.01}$ and the 3-parameter scaling law $L \propto \kappa\ T^{0.1}$,
which is more consistent with the logistic-growth model than with
classical diffusion. The findings suggest long-range correlation 
lengths in the FD-SOC system that operate in the vicinity of a critical state, 
which could be used for predictions of individual extreme events. 
We find also that eruptive flares (with accompanying CMEs),
have larger volumes $V$, longer flare durations $T$, higher EUV and 
soft X-ray fluxes, and somewhat larger diffusion coefficients $\kappa$
than confined flares (without CMEs). 
\end{abstract}


\section{INTRODUCTION}

Space and time scales are the most fundamental parameters in any
physical system. Correlations and scaling laws between spatial and
temporal parameters reveal the characteristics of various physical
transport processes, such as motion in gravitational or electromagnetic
fields, random walk processes, diffusion, conduction, turbulence,
percolation, branching theory, network systems, etc. 
Here we focus on transport processes in solar flares, which
to the best of our knowledge, are governed by magneto-hydrodynamic
(MHD) processes and kinetic particle physics in coronal plasmas.
However, kinetic theory and non-ideal MHD involve microscopic 
processes that exhibit a high level of complexity due to the large 
number of degrees of freedom, which is difficult or impossible to 
characterize with macroscopic quantities.

A relatively new method to understand the complexity of nonlinear
energy dissipation processes is the concept of {\sl self-organized
criticality (SOC)}, pioneered by Bak et al.~(1987; 1988), applied
to solar flares by Lu and Hamilton (1991), and reviewed in a number
of recent textbooks (Aschwanden 2011, 2013; Pruessner 2012). 
SOC can be considered as a basic physical mechanism, 
universally occurring in systems with many
coupled degrees of freedom in the limit of infinitesimal external
forcing. Such nonlinear systems are driven in the vicinity of a
critical state, which is maintained by a self-organizing feedback 
mechanism in a robust manner that does not need any fine-tuning of
physical parameters.
The macroscopic evolution of 
an instability in a SOC system, triggered by a random disturbance
that exceeds the local threshold of the instability, is governed by 
nearest-neighbor 
interactions on a microscopic level, which emerge and evolve over
macroscopic scales in spatial volumes that exhibit long-range
correlation lengths. These long-range correlation lengths demarcate zones with 
coherent deviations from the critical state, and thus indicate that
nature self-organizes to the {\sl vicinity of a critical state},
rather than being exactly at the edge of a critical state, which 
can be probed with correlation functions $C({\bf r}, t)$ (Jensen 1988,
Section 2.4),
\begin{equation}
	C({\bf r}, t) = \langle B({\bf r}_0, t_0)
	B({\bf r}_0 + {\bf r}, t_0 + t)\rangle_{{\bf r}_0,t_0} -
	\langle B({\bf r}_0, t_0 ) \rangle^2 \ ,
\end{equation}
where $B({\bf r}, t)$ is a physical observable in a SOC system,
such as, for instance, the non-potential magnetic field 
component in the solar corona. Nonlinear energy dissipation events 
in SOC systems, also called {\sl avalanches} in SOC jargon, thus evolve 
over macroscopic length scales $L$ that show some degree of correlation, 
such as zones with stressed or twisted magnetic fields (that are 
misaligned by some angle to the potential field), along which
avalanches can propagate and evolve. The spatial
scales of solar flares thus convey information that is equivalent to 
the correlations length. The correlation length can reveal
macroscopic zones with coherent deviations from criticality and can
be used to predict event sizes. For instance, free energy computed
from the non-potential magnetic field near the neutral line is used
to forecast flares, coronal mass ejections (CMEs), and filament
eruptions (e.g., Schrijver 2007, Falconer et al.~2011, 2012). 
Statistics of spatial 
scales $L$ of solar flares, which can be expressed as {\sl probability 
distribution functions (PDFs)} or {\sl occurrence frequency 
distributions} $N(L)$, exhibit ubiquitous powerlaw functions that are 
the hallmark of SOC systems.  Since SOC systems evolve into the vicinity 
of a critical state, where long-range correlation lengths exist (Jensen 1988, 
Section 2.4), small disturbances can trigger small or large avalanches,
which do not depend on any particular physical mechanism, such as, for
instance, magnetic reconnection processes that produce solar flares.

In this study we investigate the universal behavior of spatial and
temporal scales in solar flares, which are related to each other
in terms of a classical (or an anomalous) diffusion process, governed 
by the statistics of a random walk process (Aschwanden 2012b). 
We show that the statistical
distributions and correlations of spatial and temporal scales in
solar flares are independent of other physical parameters.
In contrast, other observables, such as photon fluxes do depend on 
physical parameters and are not universal for SOC processes. In
solar (and astrophysical) observations, photon fluxes are observed
at various wavelengths, which depend on physical parameters such as 
the electron temperature, density, or pressure, and thus scaling laws,
correlations, and size distributions of observed photon fluxes 
are wavelength-dependent, and not universal, which will be modeled 
in Paper II. 

\section{THEORETICAL DEFINITIONS}

The basic theory to interpret our statistical measurements of solar
flare parameters is the {\sl fractal-diffusive avalanche
model of slowly-driven self-organized criticality systems (FD-SOC)},
described and tested with cellular automaton simulations in
Aschwanden (2012a), and applied to solar flares in Aschwanden (2012b).
This theoretical model consists of four components: (1) scaling
laws of Euclidean spatial parameters (in terms of the scale-free
probability conjecture, Section 2.1); (2) the fractal geometry of the 
internal structure of SOC avalanches (Section 2.2); (3) the 
spatio-temporal evolution of SOC avalanches (in terms of 
fractal-diffusive transport, Section 2.3); and (4) physical scaling 
laws between observables and geometric SOC parameters (such as the 
flux-volume relationship; Paper II). The first three
parts have universal validity in the sense that they can be derived
from physics-free, pure mathematical and statistical probabilities,
while the fourth component depends on the specific physical mechanism
that controls the dynamics of a SOC avalanche and thus the observables. 
In this study we involve multi-wavelength observations of solar flares, 
which may exhibit different scaling laws depending on the wavelengths,
according to the involved physical mechanism. In this Paper I 
we analyze spatial and temporal parameters, which can be described
by universal scaling laws that govern the first three components of our
FD-SOC model, while we derive and test the scaling laws that concern 
the physics-based (fourth) component of the FD-SOC model in Paper II.

\subsection{The Scale-Free Probability Conjecture}

Typically, probability distribution functions (PDF) like 
Gaussian, binomial, Poisson, exponential, or powerlaw functions
are adopted to describe likelihoods for counting or flux measurement
scenarios. Most probability distributions in 
{\sl self-organized criticality (SOC)} systems, which produce 
nonlinear dissipation events (also called {\sl avalanches}), 
exhibit powerlaw functions over some parameter range of length scales 
and time scales. Our approach to understand the powerlaw nature of SOC 
systems is the so-called {\sl scale-free probability conjecture}, 
which is based purely on a statistical argument,
in analogy to Gaussian distributions that can be derived from 
binomial statistics.

In SOC models (Bak et al.~1987), which produce
avalanches in a multiplicative manner over some time interval of coherent
growth, the statistics is fundamentally different from the random additivity of
incoherent processes with Gaussian statistics. The classical SOC paradigm
is a sandpile with a critical slope that is self-adjusting (or self-organizing)
after each avalanche. However, the local slopes are not exactly at the
critical threshold, but show some small irregular deviations, which 
are called {\sl coarse-granularity} or {\sl micro-roughness}.
These deviations in the vicinity of the critical threshold have spatial
long-range correlation lengths that enable chain reactions over large distances
once a local instability sets in. Thus, a SOC system has the ability to
produce avalanches over a large range of sizes $S$, which sometimes
exceeds several orders of magnitude. We argue that 
the statistical probability distribution function of an avalanche with
length scale $L$ in any Euclidean space with dimension 
$d$\footnote{In the previous study we used the notation 
$S=1,2,3$ for the Euclidean spatial dimension (Aschwanden 2012a), 
but rename it here to $d=1,2,3$, to be more consistent with literature
and to avoid confusion with the parameter $S$ that is generally used 
to characterize the size of a SOC avalanche (e.g., Bak et al.~1987; 
Pruessner 2012).}
($d=1,2,3$), is (Aschwanden 2012a), 
\begin{equation}
			N(L) \propto L^{-d} \ ,
\end{equation}
which represents a powerlaw distribution function.
This simple statistical probability argument is independent of the
physical process that drives a SOC avalanche, such as gravity in the
case of sandpiles or magnetic reconnection in the case of solar flares,
and thus has universal validity for SOC systems.

Other size distributions of geometric or spatial parameters that can 
directly be derived from Eq.~(2) are the Euclidean 2-D avalanche
area $A$ and the 3-D avalanche volume $V$. 
If we simply define the Euclidean area $A$ in terms of the squared length
scale, i.e., $A \propto L^2$, we obtain the area size
distribution $N(A)$ by substitution, 
\begin{equation}
        N(A) dA \propto N[L(A)] \left| {dL \over dA} \right| dA
        \propto A^{-(1+d)/2} dA \ ,
\end{equation}
yielding $N(A) \propto A^{-2}$ for 3-D phenomena ($d=3)$.
 
Similarly we define the Euclidean volume, i.e., $V \propto L^d$,
which can be substituted to obtain the volume size distribution $N(V)$,
\begin{equation}
        N(V) dV \propto N[L(V)] \left| {dL \over dV} \right| dV
        \propto V^{-(1+2/d)} dV \ ,
\end{equation}
yielding $N(V) \propto V^{-5/3}$ for 3-D phenomena ($d=3)$.

Thus, we have universal predictions for the size distributions of
length scales, areas, and volumes of SOC avalanches that depend only
on the Euclidean space dimension $d$, which will be tested in
Section 3 with multi-wavelength images of solar flares. 

\subsection{Fractal Geometry}

The complex topology of a SOC avalanche at a given instant of time 
can be described by a fractal dimension (Bak and Chen 1989), 
for instance with the Hausdorff dimension $D_d$ (in Euclidean space 
dimension $d=1,2,3$),
\begin{equation}
	V_d = L^{D_d} \ .
\end{equation}
An example of several time instances $t$ of a SOC avalanche with 
determination of the Hausdorff dimension $D_d(t)$ as a function of
time is shown in Figs.~2 and 3 of Aschwanden (2012a). 
The mean value of the minimum
$D_{d,min}\approx 1$ and maximum dimension $D_{d,max} \lapprox d$ 
is a good approximation of the average fractal dimension $D_d$ 
of an avalanche, for each Euclidean dimension $d=1,2,3$, i.e.,
\begin{equation}
	D_d \approx {D_{d,min} + D_{d,max} \over 2} = { 1 + d \over 2} \ .
\end{equation}
So, we expect an average fractal dimension of 
$D_3=2.0$ for 3-D avalanches. 

\subsection{Fractal Diffusion in SOC Systems}

The third component of the FD-SOC model concerns the spatio-temporal
evolution of SOC avalanches, which can be characterized by the
generalized relationship for diffusion (Aschwanden 2012a),
\begin{equation}
	r(t) = \kappa (t-t_k)^{\beta/2} \ ,
\end{equation}
where $r(t)$ represents the evolution of the flare radius $r(t)$,
$t_k$ the start time of the diffusion process, 
$\kappa$ is the diffusion coefficient, and $\beta$ is the diffusion
(or spreading) 
powerlaw exponent. The case of $\beta=1$ corresponds to classical
diffusion or random walk, while the regimes of $0 < \beta < 1$
corresponds to sub-diffusion, and $1 < \beta < 2$ to super-diffusion
or hyper-diffusion (also called L\'evy flights by Mandelbrot), both 
representing anomalous diffusion processes. The lower limit of the
diffusion powerlaw exponent $\beta=0$ can be identified with a
logistic growth process (Aschwanden 2012a), while the upper limit
of $\beta=2$ corresponds to linear growth. Values of $\beta > 2$
would indicate nonlinear growth characteristics. The various
evolutions are pictured in Fig.~1. 

Since most instabilities, including SOC avalanches, start with
an exponential growth phase, and saturate at a maximum level,
it is useful to describe the evolutionary time profile with a
{\sl logistic equation}. The combined evolution of a logistic 
growth phase and and a diffusive saturation phase can be analytically 
expressed with a smooth transition inbetween (Aschwanden 2012b;
Section 2.5),
\begin{equation}
        r(t) = \quad
        \left\{
        \begin{array}{ll}
        r_{\infty} \left[ 1 + \exp(-{t - t_1 \over {\tau}_G}) \right]^{-1/3}
        & {\rm for} \ t \le t_1 \\
        \kappa (t-t_k)^{\beta/2}
        & {\rm for} \ t > t_1
        \end{array}
        \right.
\end{equation}
where $r_{\infty}$ represents the final spatial scale of $r(t)$ in
the case of strict logistic growth ($\beta=0$), and $\tau_G$
respresents the e-folding growth time. The time evolution $r(t)$
is shown in Fig.~1 for five different diffusion exponents
($\beta$=0.0, 0.5, 1.0, 1.5, 2.0). We will fit this analytical 
evolutionary function (Eq.~8) to the flare radius evolution
$r_{\lambda,th}(t)$ for different wavelengths $\lambda$ and
flux thresholds $q_{th}$ in Section 3, in order to obtain
statistics on the diffusion coefficients $\kappa$ and diffusion
exponents $\beta$.  

If the diffusion coefficient $\kappa$ is uncorrelated with the
flare radius $L=r(t=t_{end})$ or flare duration $T=(t_{end}-t_{start})$,
we expect (according to Eq.~7 for $T=[t-t_k]$) a scaling law between 
the spatial $L$ and time scale $T$,
\begin{equation}
	L \propto T^{\beta/2} \ ,
\end{equation}
which constitutes another universal scaling law, being dictated
by the statistical nature of a diffusion process, such as random
walk statistics in the case of classical diffusion ($\beta=1$).

Combining the scale-free probability distribution of length scales, 
$N(L) \propto L^{-d}$ (Eq.~2), with the universal scaling law for
time scales, $L \propto T^{\beta/2}$ (Eq.~9), we can then directly
predict the probability distribution of time scales, $N(T)$,
\begin{equation}
        N(T) dT = N(L[T]) \left| {dL \over dT} \right| dT
        \propto T^{-[1+(d-1) \beta /2]} \ dT \ ,
\end{equation}
which for 3-D phenomena ($d=3$) predicts a powerlaw slope of
$\alpha_T=1+\beta$. Thus, the powerlaw slope for time durations
will be $\alpha_T=2.0$ for classical diffusion ($\beta=1$),
or lay in the range of $\alpha_T=1,..,2$ for sub-diffusion, or
in the range of $\alpha_T=2,..,3$ for hyper-diffusion. 

\medskip
In conclusion, according to the FD-SOC model, the probability
distributions of space $L$ and time scales $T$ depend only on the 
Euclidean space dimension $d$ and diffusion exponent $\beta$. 
The internal fractal geometry of a SOC avalanche is quantified 
by the fractal dimension $D_d \approx (1+d)/2$ and has a universal 
characteristics also. In contrast, the probability distributions 
of other observable parameters, such as the flux $F_{\lambda}$
observed in a given instrumental wavelength $\lambda$,
do depend on specific physical mechanisms observed in SOC 
avalanches, which can be quantified in terms of a flux-volume 
scaling law, $F_{\lambda} \propto V^\gamma$, the fourth component
of the FD-SOC model, which we will explore in Paper II. 

\section{OBSERVATIONS, DATA ANALYSIS, AND RESULTS}

The observations and the data analysis method are similar
to the previous study of Aschwanden (2012b). The two major differences
to the previous study are that (i) we analyze a multi-wavelength data set
that includes all seven coronal filters (94, 131, 171, 193, 211, 304, 
335 \ang ) of AIA/SDO (Lemen et al.~2012; Boerner et al.~2012),
compared with the earlier single-wavelength study  
(335 \ang ), and (ii) we are using five relative flux thresholds for the
identification of flare areas, while the previous study used only two
fixed flux threshold levels. The main motivation for this extended study 
is that we want to distinguish universal (statistical) from 
wavelength-dependent (physical) scaling laws, as well as to explore 
possible systematic uncertainties in the flare area definition.
In the following subsections we discuss a number
of systematic uncertainties (such as the effects of flux 
thresholds in the flare area measurments, possible dependencies
of flux size distributions on the observed wavelengths, 
the effects of flux saturation in flare area measurements from overexposed
images, effects of coronal dimming, and uncertainties in the definition
of flare durations). Random errors due to small-number statistics and the
corresponding accuracy of powerlaw fits are discussed in Section 
7.1.7 of Aschwanden (2011).
We determined also formal errors of powerlaw fits, but do not list
them here, because larger uncertainties occur due to the deviations from
ideal powerlaws, the choice of the lower cutoffs (where incomplete 
sampling applies) and upper cutoffs (for the rare largest events). 
Instead we list empirical uncertainties of the powerlaw slopes in terms
of the means and standard deviations obtained from averaging the slopes
among different wavelengths (bottom lines in Tables 1-6) and among
different flux thresholds (last columns in Tables 1-6).

\subsection{Observations}

The dataset consists of 155 solar flares that includes all M- and X-class
flares detected with AIA/SDO during the first two
years of the SDO mission (during the time interval of 2010 May 13 and
2012 March 31). All AIA images have a cadence of $\Delta t=12$ s and a
pixel size of $\Delta x=0.6\arcsec \approx 435$ km, which corresponds
to a spatial resolution of $2.5\Delta x = 1.5\arcsec \approx 1100$ km.
We analyzed about 100 time frames for each flare in 7 wavelengths, which
amounts to a total number of $\approx 10^5$ images or $\approx 2$ 
Terabytes of AIA data.

The data analysis for a single wavelength is described in some detail
in Section 3.2 of Aschwanden (2012b), where two fixed flux thresholds
were used at $F_{335,th} \ge 100$ and $\ge 200$ DN s$^{-1}$. Here we
use five different flux thresholds $F_{\lambda,th}$ in each of the seven 
wavelengths, which were set at $q_{th}=F_{\lambda,th}/F_{\lambda,max}
=0.01, 0.02, 0.05, 0.1$, and $0.2$ of the peak flux $F_{\lambda,max}$ 
during the flare time interval. The use of multiple thresholds
yields more robust normalized flare areas, because fuzzy fine
structures of the flare boundaries are better averaged out at different
flux levels (see Section 3.2 and Figure 3a). 
Two flare area definitions are used: 
(i) The instantaneous flare area $a_{\lambda,th}(t)$ that entails all 
pixels with a flux above the flux threshold at time $t$, and 
(ii) the time-integrated flare area $A_{\lambda,th}(<t)$ that combines 
pixels that had a flux exceeding the threshold anytime during the flare
time interval $[t_{start}, t]$. We will see that the instantaneous 
flare area $a$ has a fractal characteristics, while the time-integrated 
flare area $A$ has a near Euclidean, space-filling topology. Defining an 
equivalent circular area, we can then characterize a length scale 
$r_{\lambda,th}$ that corresponds to the radius of this circular area,
\begin{equation}
	r_{\lambda,th} = \sqrt{ A_{\lambda,th} / \pi } \ .
\end{equation}

An example of the analysis of one flare event (\#28) in one single filter 
(193 \ang ) is shown in Fig.~2, showing the flux profiles $F_{193}(t)$
and flare radius profiles $r_{193,th}(t)$ for the 5 different thresholds.
This figure can be compared with Fig.~2 in Aschwanden (2012b), where
the same event is analyzed in a different wavelength (335 \ang ) with
two different thresholds. The 193 \ang\ images shown in Fig.~2 
exhibit also episodes of flux saturation due to too long exposure times,
which however, are adjusted in subsequent images with the automated 
exposure control mechanism of AIA. The flux saturation can lead to
significant underestimates of the peak flux (measured in a single pixel), 
but the missing flux in saturated flare areas represents a small fraction 
of the total image-integrated EUV flux and inferred flare area here, 
as the continuity of the total flux profiles 
$F_{193}(t)$ and flare radius profiles $r_{193,th}(t)$ show in Fig.~2. 

The 7 temperature filters of AIA are sensitive to coronal plasma in the
temperature range of $log(T) \approx 5.2-7.7$ (Boerner et al.~2012).
We will describe the thermal response functions of these wavelength
filters in Paper II, where we perform a differential emission measure 
(DEM) analysis and determine wavelength-independent physical parameters,
such as electron densities and electron temperatures.

\subsection{Normalization of Threshold-Dependent Length Scale}

Since we measure the length scale $L_{\lambda,th}$ from the flare area 
$A=\pi r_{\lambda,th}^2$ at five different flux threshold levels
$q_{th}=0.01, 0.02, 0.05, 0.1, 0.2$ of the maximum flux $F_{\lambda,max}$,
we obtain flare areas and length scales of different sizes $L_{\lambda,th}$,
where the lowest flux threshold level of $q_{th}=0.01$ encompasses the
largest flare areas, while the highest flux threshold level of $q_{th}=0.2$
encloses the smallest flare areas, among the different threshold levels. 
In order to eliminate this dependence 
of the length scale on the flux threshold, we normalize all flare areas 
to a fixed threshold level of $q_{th}=0.05$. We find empirically the 
following relationship between the length scales $L(q_{th})$ measured
at a threshold level $q_{th}$ and a normalized length scale
$L^{norm}=L(0.05)$ defined at a fixed threshold level of $q_{th}=0.05$
(Fig.~3, top panel),
\begin{equation}
	L_\lambda(q_{th}) = L_{\lambda}(0.05) 
	\left({q_{th} \over 0.05}\right)^{-0.383} \ . 
\end{equation}
This relationship works equally well for all 7 wavelengths, as demonstrated 
with the scatterplot shown in Fig.~3 (bottom panel) between the measured 
length scales $L_\lambda(q_{th})$ and the normalized length scales 
$L_\lambda(0.05) (q_{th}/0.05)^{-0.383}$. This relationship quantifies 
that the ratio $L_{\lambda}(q_{th})/L_{\lambda}(0.05)$ of the measured length 
scales are factors of 1.85, 1.42, 1.00, 0.77, and 0.59 different at the 
various thresholds $q_{th}=$ 0.01, 0.02, 0.05, 0.1, 0.2 with respect to a
chosen fixed threshold of $q_{th}=0.05$. These scale factors can be seen in 
the vertical scale of the fitted flare expansion curves $r_{\lambda,th}$ 
in the example shown in Fig.~2 (middle panel). 
The normalized flare areas
synthesized from five different thresholds average the fuzzy contours of
flare areas better than the area obtained from a single arbitrary flux 
threshold. We apply this threshold
normalization (Eq.~12) to all parameters that are derived from a measured
length scale $L$, such as to the flare area $A$, the flare volume $V$, 
the diffusion coefficient $\kappa$, and the maximum velocity $v_{max}$. 

In principle, the normalization of length scales represents a multi-variate 
problem, but the excellent correlation of the length scale with the flux 
threshold (Eq.~12 and Fig.~3, top panel) justifies a simplified bivariate 
treatment.

\subsection{Statistics of Flare Areas}

The flare area $A$ is the most directly measured quantity among the
geometric parameters, while the length scale $L$ and volume $V$ are
deduced from the area $A$. From the 155 flares we derived a flare area 
$A_{\lambda,th}(t=t_{end})$ at the end time $t_{end}$ of each flare,
for each of the 7 wavelengths $\lambda$ and 5 flux thresholds $q_{th}$.
The end time $t_{end}$ of a flare is defined (according 
to NOAA flare catalogs) when the GOES 1-8 \ang\ flux decays to the
half value of the (preflare background-subtracted) peak flux (marked 
with a vertical dashed line near the right border of the panels shown
in Fig.~2). Histogramming the observed flare areas $A$, we find flare 
areas between $A_{min}=44$ Mm$^{2}$ and $A_{max}=19,000$ Mm$^{2}$. 
All powerlaw fits in 7 wavelengths $\lambda$ and 5 thresholds $q_{th}$ 
are shown separately in Fig.~4. Since the log-log histograms of flare
areas extend over about two decades for each subset, the powerlaw
slopes can be determined somewhat more accurately ($\approx 15\%$) 
than for length scales. The values for each wavelength and threshold 
are listed in Table 2.  The overall average is, where the error bar
represents the standard deviation from averaging the area parameter 
from different wavelengths and flux thresholds (see Table 2),
\begin{equation}
	\begin{array}{ll}
		\alpha_A=2.1\pm0.3	&{\rm Observations} \\
		\alpha_A=2		&{\rm Theory} \ ({\rm for}\ d=3)  \\
	\end{array}
\end{equation}
which also agrees with the predicted value $\alpha_A=2$ for SOC
phenomena in 3-D space (Eq.~(3); for d=3). Similarly as for 
length scales, we find for flare areas no wavelength dependence 
(Figs.~5b and 6b). There is only a very slight trend that the flare 
area depends on the flux threshold (Fig.~7b and Table 2).

Inspecting the wavelength dependence of the powerlaw distributions 
in detail (Fig.~4), we do find the largest deviations from predicted 
powerlaws for the wavelengths of 171 and 193 \ang\ at low threshold 
levels of $q_{th}=0.01-0.02$. Watching the flare movies it is obvious
that the EUV dimming caused by coronal mass ejections is best
visible in the 171 and 193 \ang\ channels, because they are most
sensitive to the bulk of the ejected plasma at an electron temperature
of $T_e \approx 1-2$ MK plasma.
The ejected plasma causes a dimming in the EUV flux (due to the 
proportionality between coronal mass column depth and optically thin 
emission) in such a way that the bright flare area is substantially 
reduced and underestimated, compared with the other unimpeded wavelengths.
Thus, it is recommendable to exclude those two EUV filters for
this type of statistical studies of spatio-temporal flare parameters.
Inquiring the standard deviations for each wavelength range in
Table 2, we find that the wavelengths of 94, 131, 304, and 
335 \ang\ have the smallest variation of the powerlaw slope
$\alpha_A$ averaged over different flux thresholds, and thus are 
all equally suitable for statistical studies.

Foreshortening effects as a function of
the center-to-limb distance could introduce a cosine-dependence if
the flare volumes would be 2-dimensional (as seen in H-$\alpha$ or
white light), but observational measurements in EUV and soft X-rays,
and volume models in terms of flare loop arcades, yield a much weaker
effect that is probably not measurable due to random variations of 
the flare area geometry, orientation, and filling factors (Aschwanden
and Aschwanden 2008a,b).

\subsection{Statistics of Length Scales}

We histogrammed the length scales $L_{\lambda,th}$ (derived from the
flare area $A_{\lambda,th}$ according to Eq.~(11)) and 
fitted a powerlaw function in the upper half of the logarithmic size 
range (similarly as shown for flare areas in Fig.~4). 
The range of length scales extends from $L_{min}=3.7$ Mm to 
$L_{max}=78$ Mm for different wavelengths (normalized to a common
flux threshold of $q_{th}=0.05$), so it does cover slightly more than 
one logarithmic decade. Since the lower
half of this logarithmic length distribution shows a rollover due to
undersampling of length scales near the flux threshold (of M1 GOES
class flares) in the selection of events, we can fit only a range of 
length scales that extends over about a half decade, leading to 
uncertainties of $\approx 20\%$ in the determination of the powerlaw
slope. We tabulate the obtained powerlaw slopes $\alpha_L$ of length
scales in Table 1, and average them in each wavelength and flux
threshold separtately. The overall average and standard deviation
of all $7 \times 5$ powerlaw fits is (see also Table 1),
\begin{equation}
	\begin{array}{ll}
		\alpha_L=3.2\pm0.7	&{\rm Observations} \\
		\alpha_L=3		&{\rm Theory} \ ({\rm for}\ d=3)  \ ,
	\end{array}
\end{equation}
which agrees well with the predicted value $\alpha_L=3$ for
SOC phenomena in 3-D space (Eq.(2); for d=3). 

Another important result is that we see no dependence on the wavelength,
as the powerlaw fits of the size distributions in Fig.~5a show,
or the powerlaw slopes $\alpha_L$ as a function of the wavelength $\lambda$
shown in Fig.~6a, and listed in Table 1, which supports our assumption of 
universality for the length scale distribution $N(L)$.
However, there is a slight trend that the powerlaw slope flattens
with increasing flux threshold, from $\alpha_L=4.0\pm0.7$ at the
1\% threshold level to $\alpha_L=2.8\pm0.3$ at the 20\% threshold level
(Fig.~7a and Table 1). Obviously, the low 1\% threshold includes more 
small-scale fluctuations in the time-integrated flare area than a 
threshold of 20\%. 

\subsection{Statistics of Flare Volumes}

Thirdly, we histogram the inferred flare volumes $V$, using the Euclidean
definition for a hemisphere with radius $L$,
\begin{equation}
	V = {2 \pi \over 3} L^3 \ ,
\end{equation}
which ranges from $V_{min}=110$ Mm$^3$ to
$V_{max}\approx 10^6$ Mm$^3$. The histograms of flare volumes
extend over about three orders of magnitude and thus yield a
smaller absolute error in the powerlaw slope ($\approx 10\%$) 
than the distribution of 
length scales and flare areas. The powerlaw fits are shown separately
for each wavelength in Fig.~5c, and the powerlaw slopes are tabulated 
and averaged in Table 3. The overall average and standard ceviation is, 
\begin{equation}
	\begin{array}{ll}
		\alpha_V=1.6\pm0.2	&{\rm Observations} \\
		\alpha_V=1.67		&{\rm Theory} \ ({\rm for}\ d=3)  \\
	\end{array}
\end{equation}
which also agrees with the predicted value $\alpha_V=5/3$ for SOC
phenomena in 3-D space (Eq.~(4); for d=3). Also here we find
no significant dependence of the powerlaw slope on the wavelength 
(Fig.~6c, Table 3) or flux threshold (Fig.~7c, Table 3).
Thus, the size distribution of volumes does not depend on the flux
threshold or observed wavelength. 

\subsection{Statistics of Flare Durations}

For the duration of the 155 analyzed flare events we use the rise time
$T=(t_{SXR}^{peak}-t_{SXR}^{start})$ of GOES flares (Fig.~1, top), 
which reflects the impulsive phase of the flare without the arbitrarily 
long cooling times in the decay phase of the flare. For the same reason,
the duration of energy release cannot easily be determined from EUV
data, because the bulk of EUV emission originates late in the postflare
phase when the flare plasma cools down to EUV temperatures. 
The energy release time 
or impulsive phase of a flare can be best defined by the duration of 
nonthermal hard X-ray emission $T=(t_{HXR}^{end}-t_{HXR}^{start})$
(Fig.~1, top). Using the Neupert effect, which states that the hard X-rays
approximately follow the time derivative of the soft X-rays
(as measured by GOES), the peak time $t_{peak}$ of the GOES light
curve approximately signals the end of the nonthermal hard X-rays 
(Fig.~1 top), which justifies that we use the rise time of the 
soft X-rays as a measure of the energy release time or nonthermal 
flare duration $T$. Since the definition of an event duration
is most critical for systematic errors and uncertainties of the
powerlaw distribution of event durations, we emphasize that the 
nonthermal flare duration (ignoring the thermal cooling phase) 
is most consistent with the duration of energy release avalanches 
in self-organized criticality systems. 

The official definition of the flare duration used in the NOAA flare 
detection algorithm is: The event starts when 4 consecutive 1-minute 
X-ray values have met
all three of the following conditions: (i) All 4 values are above the B1
threshold; (ii) All 4 values are strictly increasing; (iii) The last value
is greater than 1.4 times the value that occurred 3 minutes earlier.
The peak time is when the flux value reaches the next local maximum.
The event ends when the current flux reading returns to half of the peak
value (http://www.ngdc.noaa.gov/stp/solar/solarflares.html).

The flare occurrence frequency distribution $N(T)$ of GOES flare durations 
$T$ is shown in Fig.~5d, which exhibits an approximate powerlaw slope of 
\begin{equation}
	\begin{array}{ll}
		\alpha_T=2.1\pm0.2	&{\rm Observations} \\
		\alpha_T=2.00		&{\rm Theory} \ ({\rm for}\ d=3)  \\
	\end{array}
	\ ,
\end{equation}
where we determined a mean and standard deviation from using 10 different
trials of histogram binning. This value
agrees with the predicted value $\alpha_T=2$ for SOC
phenomena in 3D Euclidean space ($d=3$) and classical diffusion ($\beta=1$)
according to our FD-SOC model (Eq.~9). While our statistics with a
relatively small sample of 155 events is rather humble, larger statistics
with over 300,000 GOES flare events has been conducted in Aschwanden
and Freeland (2012), where a powerlaw slope of $\alpha_T=2.02\pm0.04$
was found during solar cycle minima, while steeper values are observed
during solar cycle maxima, as a consequence of the solar cycle-dependent 
flare pile-up bias effect.

\subsection{Spatio-Temporal Relationships}

After we gathered statistics of spatial $(L, A, V)$ and temporal scales $T$
separately, we turn now to spatio-temporal relationships, which we model
with a generalized diffusion equation (Eq.~7), specified by a diffusion
coefficient $\kappa$ and a diffusion exponent $\beta$. Histograms of the
parameters $\kappa$ and $\beta$ are shown in Figs.~5f and 7g (for a single 
threshold of $q_{th}=0.05$), separately sampled for each of the 7 observed 
wavelengths $\lambda$, and tabulated in Tables 4 and 5. 

The diffusion coefficient
has a range of values between $\kappa=4.2$ and $76.7$ Mm s$^{-\beta/2}$,
with an overall mean of $\kappa=29.7\pm7.7$ Mm s$^{-\beta/2}$ (Table 4),
which shows a slight wavelength dependence within a factor of $\approx 1.5$
(Fig.~6f), but no threshold dependence after normalization to a common
threshold (Fig.~7f), see Section 3.2. 

The diffusion exponent $\beta$, for which we expect a value of $\beta=1$ 
for classical diffusion, turned out to be sub-diffusive, measured as 
$\beta=0.53\pm0.27$ for the 335 \ang\ wavelength in a previous study
(Aschwanden 2012b). Here, we confirm the same result of sub-diffusity,
where we find a consistent value of $\beta=0.50\pm0.04$ at the same
wavelength of 335 \ang, although we used different flux threshold levels.
Averaging over all wavelengths we find also a similar value of
$\beta=0.35\pm0.13$ (Table 5). The histograms of the diffusion
exponents $\beta$ show consistent distributions in each of the 7 wavelengths
(Fig.~5g), and no particular dependence on the wavelength (Fig.~6g) or
flux threshold (Fig.~7g). 

In Figure 8 we show scatterplots of those pairs of parameters that have
the highest correlation coefficient ($ccc > 0.4$), along with linear
regression fits using the {\sl orthogonal reduced major axis} method
(Isobe et al.~1990), with the mean and standard deviation of the linear
slope indicated. From all possible correlations between the parameters 
$L$, $T$, $\kappa$, $\beta$, and $v_{max}$, we find that the highest 
correlation occurs between the diffusion coefficient $\kappa$ and the 
length scale $L$, with a linear regression fit of 
\begin{equation}
	\kappa(L) \propto L^{0.94\pm0.01} \ ,
\end{equation}
with a Pearson correlation coefficient of $ccc=0.96$ (Fig.~8a).
This is an unexpected new result, since we assumed an uncorrelated 
diffusion coefficient $\kappa$ before. 
A question of special interest is whether the spatio-temporal data gathered 
here are consistent with the fractal-diffusive FD-SOC model, which assumes
a scaling law of $L \propto T^{\beta/2}$ (Eq.~9). If we investigate
a scatterplot, we find indeed a scaling of $L \propto T^{0.46\pm0.03}$
(Fig.~8b), which implies, under the assumption of a constant diffusion 
coefficient $\kappa$, a diffusion exponent of $\beta=0.92\pm0.06$, 
which is consistent with classical diffusion. The correlation between
spatial and temporal time scales, however, is weak, with a Pearson
correlation coefficient of $ccc=0.42$ (Fig.~8b).

The best three-parameter correlation actually occurs for the relationship 
\begin{equation}
	L \propto \kappa \ T^{0.1} \ ,
\end{equation}
with a Pearson correlation coefficient of $ccc=0.97$ (Fig.~9), being 
valid for all 7 wavelengths (marked with different colors in Fig.~9).
Thus, the flare size scale $L$ strongly depends (almost linearly)
on the diffusion coefficient $\kappa$, but is almost uncorrelated
with the time duration $T$. If we apply the logistic model (Fig.~1
and Eq.~8) instead of the fractal-diffusive model, we indeed expect 
that the final flare size scale $L$ is proportional to the maximum
velocity $v_{max}$ (Eq.~20) and to the diffusion coefficient $\kappa$,
with no dependence on the duration $T$ in the limit of $T \gg \tau_G$, 
which is approximately consistent with Eq.~(19).

\subsection{Statistics of Maximum Velocities}

The diffusion equation (Eq.~7) implies a variable expansion speed $v(t)$,
but we can define a maximum speed $v_{max}=v(t_1)$ at the transition time
$t_1$ between nonlinear growth and the onset of diffusion (Eq.~8),
which is the inflection point of the derivative $dr/dt$. The maximum
velocity $v_{max}$, occurring at this time $t_1$ is,
\begin{equation}
	v_{max}={dr \over dt}(t=t_1)=
	{\kappa \beta \over 2} (t_1-t_k)^{\beta/2-1} 
	= {\beta \over 2} {r_1 \over (t_1-t_k)} \ .
\end{equation} 
Thus, we expect some correlation of the maximum velocity $v_{max}$
with the length scale $L$ and time scale $T$, by setting $L = r(t=t_1)=r_1$
and $T=(t_1=t_k)$ in Eq.~(20), 
\begin{equation}
	v_{max} = {\beta \over 2} {L \over T} \ .
\end{equation}
For the fractal-diffusive model with $L \propto T^{\beta/2}$
this would imply $v_{max} \propto T^{\beta/2-1} \approx T^{-1/2}$ for
classical diffusion ($\beta=1$). However, 
the scatterplots shown in Fig.~8 exhibit some weak correlations
between those parameters, with linear regression fits of 
$v_{max} \propto T^{-1.1}$ (Pearson correlation coefficient $ccc=0.43$;
Fig.~8c), which is more consistent with the logistic model, where
$L$ and $T$ are uncorrelated and $v_{max} \approx T^{-1}$ is expected.

\subsection{Fractal Geometry}

We explore now the internal geometric structure of solar flares,
which we described with a fractal Hausdorff dimension $D_d$ in our
FD-SOC model (Eq.~5). 
We measured the instantaneous flare area $a(t)$ during the flare
time interval $[t_{start} < t_i < t_{end}]$ for each of the 155 flares,
consisting of typically 100 time frames per flare, by counting the
pixels that had a flux in excess of a given flux threshold, for five 
different thresholds, i.e., $q_{th}=0.01$, 0.02, 0.1, 0.2, 0.5 of the maximum
(preflare background-subtracted) flux during the flare time interval,
and for each of the 7 coronal wavelengths. The Euclidean length scale $L$ 
was determined from the time-integrated flare area $A \propto L^2$ 
at the end time $t_{end}$ of the flare. The 2-D fractal Hausdorff 
dimension $D_2$ was then determined from the relationship 
$A \propto L^{D_2}$ according to Eq.~(5),
\begin{equation} 
	D_2(t) = {\log{A(t)} \over \log{L}} \ ,
\end{equation}
from which the 3-D fractal dimension $D_3=(4/3) D_2$ can be inferred.
An example of the fractal dimension $D_2(t)$
as a function of time during the flare \#28 (the same as shown in Fig.~2)
is plotted in Fig.~10. The time-evolution of the fractal dimension
$D_2(t)$ can be generally characterized by a rapid initial increase that
settles into a constant value after the first impulsive peak of the flare.
The example shown in Fig.~10 demonstrates that this behavior is
similar at different wavelengths and flux thresholds. The lowest flux
thresholds yield at times a slightly higher fractal dimension. We
histogram the fractal dimensions for each wavelength in Fig.~5,
averaged over all 155 flares and 5 flux thresholds, but separately
for each wavelength. Apparently there is no wavelength dependence of
the fractal dimension, as the average values show, tabulated in Table 6.
Averaging also over all wavelengths, we find a mean value of
$D_2=1.55\pm0.11$ for the 2-D area fractal dimension, and 
$D_3=2.07\pm0.26$ for the 3-D volume fractal dimension, which is
consistent with the theoretical prediction of the FD-SOC model,
with $D_2=1.5$ or $D_3=2.0$. This is a very useful result that allows
us to estimate the emitting flare volume $V=L^{D_3}$ from the
flare length scale $L$, which is relevant for quantifying the 
observed flux $F_{\lambda}$ from the emitting flare volume $V$.

Note, that the fractal dimension $D_3$ serves here mostly to estimate
the filling factor of the emitting flare plasma. We do not test
the self-similarity of the flare geometry at different spatial scales  
or spatial resolutions for a given flare at a given time here
(as measured from a set of 20 large flares in Aschwanden and 
Aschwanden 2008a,b; or from a cellular automaton avalanche 
in Aschwanden 2012a), but rather measure the self-similarity as a 
function of time during a given flare (e.g., Fig.~10), as well as 
among the entire set of analyzed flares.
The so inferred 3-D fractal dimension with a mean and standard 
deviation of $D_3=2.1\pm0.3$ is nearly invariant among different
flare events, as well as during the main flare time interval 
(from the peak time $t_{peak}$ to the end time $t_{end}$).

\section{DISCUSSION}

\subsection{Universal Scaling of Space and Time Parameters}

In this Section we discuss observational evidence of our theoretical assumption 
that the statistical probability distribution functions (PDF) of geometric 
(or space) and temporal (or time) scales of nonlinear dissipative 
(avalanche) events in SOC systems are governed by a universal scaling 
that can be derived from a pure statistical probability argument,
regardless of the physical mechanism involved in the generation
of a SOC avalanche. This issue is tackled here for the first time
with solar data sampled over a broad wavelength and temperature 
range, and thus over a wide range of physical conditions.

Let us first compare our measurements with previous mostly single-wavelength 
observations of solar flares. Table 7 offers a compilation of observed
size distributions. The most directly measured geometric quantity
is the flare area $A$, while the length $L \propto \sqrt{A}$ (Eq.~11)
and volume $V \propto L^3 \propto A^{3/2}$ (Eq.~15) are quantities that 
are generally derived
from the flare area $A$. For 18 datasets listed in Table 7, frequency
distributions for the flare area were measured
(Berghmans et al.~1998; Aschwanden et al.~2000; Aschwanden and Parnell 2002;
and in this work), with a mean powerlaw slope
of $\alpha_A=2.2\pm0.2$, which is consistent with the theoretical
prediction $\alpha_A=2$ (for Euclidean space dimension $d=3$). Note, 
that about half of these 18 datasets
were measured from nanoflares, while the other half is measured from
giant flares (of GOES M- and X-class). So, we find neither a
dependence of the flare area distributions $N(A) \propto A^{-\alpha_A}$
on the wavelength $\lambda$, nor on the magnitude of the flare, 
which is also true for the other derived distributions, of length
scales $N_L \propto L^{-\alpha_L}$, and volumes $N_V \propto V^{-\alpha_V}$,
and thus we can conclude that the size distributions of geometric
parameters indeed appear to be universal, at least for the phenomenon
of solar flares, as the compilation with a number of different
instruments and wavelength ranges demonstrates in Table 7. 
Additional support for the universality of the powerlaw slope of $\alpha_L
\approx 3$ is also given by the size distributions of lunar craters,
asteroids, and Saturn ring particles, which have been associated with
SOC mechanisms (Aschwanden 2013). 

Size distributions of time scales $T$ were measured for a larger
number of datasets (Crosby et al.~1993, 1998;
Lu et al.~1993; Lee et al.~1993; Bromund et al.~1995; Berghmans
et al.~998; Aschwanden et al.~2000; Veronig et al.~2002; 
McIntosh and Gurman 2005; Su et al.~2006; Yashiro et al.~2006;
Christe et al.~2008; Nishizuka et al.~2009; Aschwanden 2011, 2012b;
Aschwanden and Freeland 2012; and in this work), as listed in Table 7.
If we exclude those measurements that fitted a broken powerlaw
(Crosby et al.~1998; Georgoulis et al.~2001; Su et al.~2006), 
a powerlaw with an exponential fall-off (McIntosh and Gurman 2005), 
or those without preflare background subtraction 
(Veronig et al.~2002; Yashiro et al.~2006), we have 20 independently
measured datasets with a mean powerlaw slope of $\alpha_T=2.3 \pm 0.4$,
which is consistent with the theoretical prediction $\alpha_T=2$. 
These time scale distributions $N(T) \propto T^{-\alpha_T}$ have been
measured from photons with different emission mechanisms, such as
nonthermal bremsstrahlung at hard X-ray energies of $>25$ keV, or
thermal line emission in soft X-rays ($1-8$ \ang) and EUV wavelengths 
($94-335$ \ang). The fact that we measure the same powerlaw slope for
different physical emission mechanisms, which is moreover consistent
with the probability distribution $N(T) \propto T^{-2}$ predicted by
the FD-SOC model, strongly supports the interpretation that the time
scale distribution has universal validity also, depending only on
the Euclidean dimension ($d=3$) of the length scale distribution
$N(L) \propto L^{-d}$ and on the random walk statistics 
$L \propto T^{1/2}$ that defines the spatio-temporal transport
process. 

\subsection{Finite-Size Effects of Active Regions}

Solar flares do not occur uniformly distributed over the entire solar
surface, but are generated in active regions only, where strong 
magnetic fields emerge, generated by the internal solar dynamo.
Every active region thus represents a finite-size SOC system, limiting
also the maximum size of solar flares. In our measured distribution
of flare length scales $L$, defined from the radius of a circular
area that is equivalent to the observed flare area $A = \pi L^2$,
we find a range of approximately $L \approx 5-100$ Mm
(Figs.~5a and 9), which
corresponds to diameters in the range of $2L=10-200$ Mm for circular
flare areas. If the flare area is highly elongated, the major axis
can be factor of $\lapprox 5$ longer than the minor axis. 
One of the largest observed
flares is the Bastille-Day flare, which has an arcade length of
$l \approx 180$ Mm and a width of $w \approx 35$ Mm
(Aschwanden and Alexander 2001), which indeed has an elongation
factor of $l/w \approx 5$. The largest flare we observed in our
selection has a diameter of $2L \lapprox 200$ Mm, which amounts to about
a quarter of the solar radius, corresponding to the largest active 
regions seen on the Sun (e.g., Tang et al.~1984). Thus, the size
of flares can occupy almost entire active regions, and consequently
we would expect finite-size effects that limit the upper boundary
of the size distribution, and potentially could produce a bump 
at the upper cutoff of the size distribution (e.g., as it happens
for auroral blobs; Lui et al.~2000).
Such an effect is not evident in our data (Fig.~5a) and has not been 
reported somewhere else, which suggests that there is no significant 
number of giant flares that reach the full size of active regions. 
Moreover, the largest flares occur in active regions of different 
sizes, which smears out the upper cutoffs in the combined size 
distributions of flares. This is an important detail that effects the 
forecasting of the most extreme (space weather) events.

\subsection{Solar Cycle Effects}

Although the size distribution of flare durations can be considered
as universal, there is a non-universal effect that is caused by the
solar cycle, which controls the variation of the flare rate over several 
orders of magnitude, from the minimum to the maximum of the solar 
magnetic 11-year cycle. This effect was most clearly measured in a 
large sample of
over 300,000 solar flares detected with GOES during the last 37 years
(Aschwanden and Freeland 2012). The flare rate during the solar maximum
increases to a level that does not warrant the separation of time scales
anymore, i.e., the waiting times (i.e., the time intervals between
subsequent flares) are not anylonger much larger than the flare durations, and
thus subsequent flare events start to pile-up, which leads to an
under-estimate of the true duration of large flares (since the previous
flare has to end before the next flare starts in most automated
flare detection algorithms). A consequence of such a fast-driven system 
is a steepening of the size distribution of flare durations during times
of high flare rates. The study of Aschwanden and Freeland (2012) has
demonstrated that the powerlaw slope of flare durations amounts to 
a lowest value of $\alpha_T=2.02\pm0.04$ in years the near solar minimum, 
close to the theoretical expectation of $\alpha_T=2$, but to a
much steeper value during solar maxima, approaching values as high as 
$\alpha_T \lapprox 5$ during the most flare-active years.
Nevertheless, this is mostly a technical detection problem and could
in principle be eliminated if the true flare duration would be measured
during times of high flare rates by modeling the time profiles in the
decay phase beyond the overlap with the next flare. So, it is a measurement
problem only, rather than a non-universal property of the size distribution
of event durations in solar flares. 

\subsection{Effects of CMEs and EUV Dimming}

Investigating the measured flare area size distributions $N(A)$ in 7
different wavelengths and in 5 different flux thresholds (Figs.~4, 6b, 7b),
as well as the powerlaw slopes listed in Tables 1, 2, and 3 for 
lengths $L$, areas $A$, and volumes $V$, it becomes apparent that the
largest deviations from the averages occur in the wavelengths of
171 and 193 \ang , especially at the lowest flux thresholds
of $q_{th}=0.01$ and 0.02. Inspecting the flare movies in those
wavelengths it becomes clear that these wavelengths show most prominently
the EUV dimming that is caused after the ``evacuation'' of a coronal mass
ejection (CME), leading to an underestimate of the flare-related
EUV emission. Since up to 84\% of CMEs exhibit
EUV dimming (Bewsher et al. 2008), this is quite common, and shows up
as a double-peaked size distribution of flare areas in the wavelengths
of 171 and 193 \ang\ (Fig.~4). This is an additional non-universal
effect that is idiosyncratic to solar flares associated with CMEs.
An analogy to Bak's sandpile paradigm of SOC systems would be huge 
sand avalanches that cause an additional dust storm which occults part
of the avalanche, and thus would affect the accuracy in the measurement
of sand avalanche sizes. In order to obtain accurate size distributions
of solar flares, we thus recommend to avoid the 171 and 193 \ang\
wavelengths, at least not for large eruptive flares, where the percentage
of CMEs and EUV dimming is highest (Andrews 2003; Yashiro et al.~2005; 
Wang and Zhang 2007; Cheng et al.~2010). 

\subsection{Eruptive Versus Confined Flares}

The spatio-temporal relationship $L(T) = \kappa T^{\beta/2}$ (Eq.~7) 
is quantified
by a generalized diffusion process in our FD-SOC framework. Interestingly,
the diffusion coefficients $\kappa$ and diffusion (or spreading) exponents
$\beta$ appear not to be constant for the analyzed sample of large flares, 
but show a rather large variation from sub-diffusion to classical diffusion
(statistical random walk), while a few events are even in the hyper-diffusive 
regime. In order to gain more physical insight into this behavior, we
consulted a flare classification catalog of the same 155 analyzed events
that was established by Zhang and Liu (2012). In this catalog, GOES
M- and X-class flares, analyzed from SDO, SOHO, and STEREO data, were classified
based on their association with CMEs, or lack of CMEs, into eruptive versus
confined flares. An eruptive flare is defined strictly as a flare associated 
with a coronal mass ejection (CME) seen in coronagraph images, while a 
confined flare exhibits no noticable CME. The verification of the flare type 
is carried out straightforwardly through direct comparison with 
concurrent coronagraph images, and coronal EUV images as well. For each of 
the major SDO flares, the SOHO/LASCO images were carefully checked
for a possible association with a CME, and potential disk signatures
were also checked in AIA/SDO images, including flare brightenings, EUV dimming,
and post-eruption arcades. In some cases, when necessary, STEREO/COR2 
coronagraph images and EUVI coronal images were checked in addition for 
further verification. Therefore, the classification of flares into eruptive 
and confined types is believed to be unambiguous. Previous studies have 
shown that about 90\% of X-class flares and about 50\% of M-class 
flares are associated with CMEs; in other words, 10\% of X-class flares 
and 50\% of M-class flares are confined (Andrews 2003; Yashiro et al.~2005; 
Wang and Zhang 2007; Cheng et al.~2010). From the 155 events analyzed in this
study, we have an overlap of 125 events with the catalog of Zhang and Liu
(2012), for which we find 59 eruptive (47\%) and 66 confined (53\%) flare 
events.

In Fig.~11 we show histograms of the various measured parameters 
$(L, V, T_{AIA}, F_{GOES}, T_{GOES}, v_{max}$, $F_{AIA}, \kappa, \beta)$
at a wavelength of 335 \ang ,
separated into the two groups of eruptive flares (Fig.~11, white histograms)
and confined flares (Fig.~11, grey histograms). 
Interestingly, we find
significant differences in the parameter distributions for flare volumes $V$,
GOES durations $T_{GOES}$, EUV fluxes $F_{AIA}$, and in the diffusion
coefficient $\kappa$. Eruptive flares have larger volumes $V$, longer
soft X-ray durations $T_{GOES}$, larger EUV fluxes $F_{AIA}$, and somewhat
larger diffusion coefficients $\kappa$. The differences between eruptive
and confined flares are very similar in all 7 analyzed wavelengths (Fig.~11
shows only one of them, i.e., 335 \ang ).
Some of these differences are expected
according to some theoretical CME models. For instance, eruptive events
are expected to have a more expansive dynamics, and thus a larger
diffusion coefficient than the confined ones, which are more stationary 
flare events, occurring within a closed-field magnetic configuration. 
Eruptive events seem also to release more magnetic energy than confined
flares, leading to statistically larger volumes $V$ and fluxes
($F_{SXR}, F_{AIA}$). 

\subsection{Diffusion Spreading Coefficient and Long-Range Correlations}

We described the results of the spatio-temporal relationships in
Section 3.7 and showed the correlations among different parameters
in Fig.~8. The most remarkable correlation, which gives us the best
hint of a hidden scaling law is the almost linear relationship 
$\kappa \propto L^{0.94\pm0.01}$ (Eq.~18) between length scales $L$ and 
diffusion coefficients $\kappa$ (Fig.~8, panel a). The uncertainty
in the linear regression coefficient amounts to only $\approx 1\%$.
Including also a possible dependence on the time duration $T$, which
is expected from random-walk transport processes, we find an equally
tight three-parameter correlation $L \propto \kappa T^{0.1}$ (Fig.~9).
Since the spatio-temporal evolution fits $L(t) = \kappa T^{\beta/2}$
(e.g., Fig.~2, middle panel) are most consistent with a constant
diffusion coefficient per flare event and subdiffusive spreading 
exponents of $\beta=0.35\pm0.13$ (Table 5), the statistical correlation
among many flare events is dominated by the diffusion coefficient
$\kappa$ rather than by the spreading exponent $\beta$. Therefore,
flare events that have a small diffusion coefficient $\kappa$ expand
to relatively small spatial scales $L$, while other events with large
local diffusion coefficients $\kappa$ produce the largest flare sizes.
The diffusion coefficient $\kappa$ is therefore the most critical
and most controlling parameter of the flare size $L$. 

In SOC systems,
a critical threshold determines whether a nonlinear dissipation event
starts or not, while the size of a SOC avalanche is mostly controlled
by the spatio-temporal correlation characteristics (Eq.~1) of the
SOC system. Locations with long-range correlation lengths produce predictably
larger SOC avalanches. Long-range correlations become important
when significant deviations from the critical threshold extend over
a larger spatial area or volume. This is another way of saying that
a SOC system self-organizes to the {\sl vicinity of a critical state},
rather than being exactly at the critical threshold. We illustrate
this behavior with the sandpile analogy in Fig.~12 (left). The slope 
of a critical sandpile shows coarse-graininess and significant deviations
from the critical slope, especially in places that have been eroded
from previous large avalanches, which can be quantified by a 
cross-correlation function $C(x,t)$ that exhibits significant
long-range correlation lengths in extended deviant locations (Fig.~12, left).
Subsequent avalanches triggered in those deviant sub-critical locations 
may evolve predictably into relatively large SOC avalanches.
In the situation of solar flares, the non-potential magnetic field
component $B_{NP}$ plays the same critical role as the repose angle
or slope of the sandpile. Extended locations with coherent non-potential
fields (Fig.~12, right) exhibit long-range correlation lengths and are more 
likely to product larger avalanches subsequently.
Observations have shown that the length of the neutral line segment
with highly sheared and stressed magnetic fields is a good predictor
of flares (e.g., Schrijver 2007, Falconer et al.~2011, 2012). 
The curl of the non-potential field corresponds to the electric
current density ${\bf j}/4\pi = \nabla \times {\bf B}_{NP}$ that
drives the dissipation of magnetic energy during flares, once it
exceeds a critical threshold $|{\bf j}| \ge j_{\rm crit}$
(e.g., Vassiliadis et al.~1998; Isliker et al.~1998a, 2000,
2001; Galsgaard 1996; Morales and Charbonneau 2008a, b, 2009).
It is therefore important to develop reliable {\sl nonlinear force-free
magnetic field (NLFFF)} models that measure the spatial and temporal
distribution of the current density ${\bf j}({\bf x},t)$ in active
regions, which map out long-range correlation lengths that are susceptible
to areas where large flares are most likely to occur, a task that
is at the heart of forecasting centers that aim to predict large and 
extreme (space weather) events. 
SOC systems may provide in this way not only statistical 
overall probabilities of events, but also probabilities for sizes 
and start times of individual single events. In order to achieve this,
future studies should focus on characterizing relationships between
current distributions  ${\bf j}({\bf x}, t)$ derived from NLFFF
models and the diffusion (or spreading) exponent $\kappa$ that can be
measured by fitting the generalized diffusion equation
$r(t)=\kappa (t-t_k)^{\beta/2}$ (Eq.~7) as performed in this study.  
 
\section{CONCLUSIONS}

In this study we analyzed the spatio-temporal evolution of a statistically
complete data set (of GOES M- and X-class flares observed with AIA/SDO
during the first two years of the mission), from which we measure
geometric parameters (lengths, areas, volumes, fractal area dimensions), 
temporal parameters (flare durations), and spatio-temporal parameters 
(diffusion coefficient, diffusion or spreading powerlaw exponent, 
maximum diffusion velocity). In this study we expanded a previous
single-wavelength (335 \ang ) study (Aschwanden 2012b) to the entire range 
of 7 coronal wavelengths of AIA (94, 131, 171, 193, 211, 304, 335 \ang ),
and generalize the flare area definition by including 5 different flux 
threshold levels. In addition we make use of a flare catalog that provides
a classification into eruptive and confined flares (Zhang and Liu 2012).
The statistical distributions obtained in this study are compared with
the predictions from the theory of the fractal-diffusive avalanche model 
of a slowly-driven self-organized criticality system (Aschwanden 2012a).
The major conclusions of this statistical study are:

\begin{enumerate}
\item{The spatio-temporal evolution of flares can be satisfactorily
fitted in all events and in all coronal wavelengths with the generalized 
diffusion model, which includes an initial acceleration phase with
exponential growth, followed by a deceleration phase with diffusive
random-walk characteristics $r(t)=\kappa (t-t_k)^{\beta/2}$.
We find a mean diffusion coefficient of $\kappa=30\pm8$ Mm s$^{\beta/2}$
and a mean diffusion (or spreading) exponent of $\beta=0.35\pm0.13$,
which is in the sub-diffusive regime (likely to be caused by the
anisotropy of the magnetic field in the flare regions).}

\item{The strongest correlation among spatio-temporal parameters
is found between the flare length scale $L$ and the diffusion 
coefficient $\kappa$, i.e., $\kappa \propto L^{0.94\pm0.01}$, 
which combined with the weaker correlation with time scales $T$,
i.e., $L \propto T^{0.46\pm0.03}$, being consistent with classical random
walk diffusion, leads to the 3-parameter scaling law 
$L \approx \kappa \ T^{0.1}$. We interpret the strong correlation
between flare size $L$ and the diffusion coefficient $\kappa$ in
terms of long-range correlation lengths in the vicinity of a self-organized
criticality state, predicting a physical relationship between the
local diffusion coefficient $\kappa$ and the non-potential magnetic
field component $B_{NP}$, which can be used for forecasting of flare
sizes $L$ by analyzing the pre-flare non-potential magnetic field.}

\item{The observed size distributions of geometric ($L, A, V, D_d$) and
temporal parameters ($T$) are consistent with the theoretical 
predictions of the FD-SOC model, i.e., $N(L) \propto L^{-3}$,
$N(A) \propto A^{-2}$, $N(V) \propto V^{-5/3}$, $N(T) \propto T^{-2}$,
$D_2=3/2$, and $D_3=2$ for a Euclidean dimension of $d=3$, and thus 
confirm the scale-free probability conjecture and the fractal
random walk diffusivity assumption of the FD-SOC model. While this
agreement of the powerlaw slopes between observations and theoretical
predictions has been shown in one single wavelength earlier
(Aschwanden 2012b), we demonstrate the same agreement here for all
coronal wavelengths observed with AIA and thus conclude that the
powerlaw slopes of the geometric and spatial size distributions 
predicted by the FD-SOC model have universal validity, independent
of other physical parameters such as electron densities, electron
temperatures, or wavelengths.} 

\item{The size range of the analyzed flares, which represent the
largest events (GOES M- and X-class) occurring on the Sun, are found
to cover a (diameter) range of $2L \approx 10-200$ Mm, which is close
to the size of the largest active regions on the Sun. However, we do
not find any evidence for finite-size effects of SOC systems that
would modify the upper cutoff of the powerlaw-like size distributions.}

\item{Non-universal effects that affect the size distribution of solar
flares are found to result from CMEs and the associated EUV dimming,
which leads to underestimates of the EUV flux in wavelengths of
171 and 193 \ang , producing a double-peaked size distribution.
Also the solar cycle can affect the powerlaw slope of measured size
distributions of flare durations $T$, which are systematically
underestimated due to flare pile-up during episodes of high flaring
rates, as they occur during the solar maximum (Aschwanden and Freeland
2012).}

\item{Eruptive flares (associated with a CME) have larger volumes $V$, 
longer soft X-ray durations $T_{GOES}$, larger EUV fluxes $F_{AIA}$, 
and somewhat larger diffusion coefficients $\kappa$ than confined
flares (without accompanying CME). Eruptive events seem to release 
more magnetic energy than confined flares, leading to statistically 
larger volumes $V$ and fluxes ($F_{SXR}, F_{AIA}$).}
\end{enumerate}

The major satisfaction of these results is our increasing understanding
of the universality of self-organized criticality systems. We understand
now the perplexing question why we observe powerlaw distributions in
nature in the first place, in contrast to Gaussian distributions.
Gaussian or exponential distributions indicate incoherent processes
that cannot explain extreme (``king-dragon'' or ``black-swan'') events
in the ``fat tail'' of probability distribution functions, while
powerlaw distributions are naturally obtained from the scale-free
probability conjecture (Eq.~2), which occur due to the coherent
nature of multiplicative chain reactions, requiring long-range
correlation lengths (Eq.~1) in the vicinity of a self-critical state.
A logical next step is the study of relationships between long-range
correlation of SOC-critical parameters, such as the nonpotential
magnetic field ${\bf B}_{NP}({\bf x},t)$ in solar flaring regions, and
the magnitude or size $L$ of occurring solar flares, which is found 
to be directly proportional to the diffusion spreading exponent $\kappa$.
Besides the universality of scaling laws for space and time parameters
in SOC systems, there are also non-universal scaling laws that depend
directly on the relationship between observables (such as fluxes in
different wavelengths) and the geometric avalanche (or flare) volume,
which will be pursued in Paper II. In future studies, the statistics
of flare events should be extended from the giant flares analyzed here,
to the smaller microflares and nanoflares, in order to improve the
accuracy of the measured powerlaw slopes (over several decades)
and to test over what maximum range of parameter space the same 
physical scaling laws are valid.

\acknowledgements
The author acknowledges helpful discussions and software support
of the AIA/SDO team, as well as constructive comments by 
anonymous referees. This work has benefitted from fruitful
discussions with Henrik Jensen, Nicholas Watkins, J\"urgen Kurths,
and by the {\sl International Space Science Institute (ISSI)} 
at Bern Switzerland, which hosted and supported a workshop on 
{\sl Self-Organized Criticality and Turbulence} during October 
15-19, 2012. This work was partially supported by NASA contract 
NNX11A099G ``Self-organized criticality in solar physics''
and NASA contract NNG04EA00C of the SDO/AIA instrument to LMSAL.

\clearpage

%%%%%%%%%%%%%%%%%%%%%%%%%%%%%%%%%%%%%%%%%%%%%%%%%%%%%%%%%%%%%%%%%%%%%%%

%__________________________TABLE 1________________________
\begin{deluxetable}{rrrrrrr}
\tabletypesize{\normalsize}
\tabletypesize{\footnotesize}
\tablecaption{Statistics of powerlaw slopes $\alpha_L$ 
of length scales L[Mm] of
155 solar flares, tabulated in 7 AIA wavelengths and for 5 different
flux thresholds. The theoretical prediction
of the FD-SOC model is $\alpha_L=3$.}
\tablewidth{0pt}
\tablehead{
\colhead{Wavelength}&
\colhead{Threshold}&
\colhead{Threshold}&
\colhead{Threshold}&
\colhead{Threshold}&
\colhead{Threshold}&
\colhead{All}\\
\colhead{[$\ang$]}&
\colhead{1\%}&
\colhead{2\%}&
\colhead{5\%}&
\colhead{10\%}&
\colhead{20\%}&
\colhead{}}
\startdata
      94 &  3.4  &  3.6 &  2.2 &  3.2 &  2.9 &  3.1$\pm$ 0.6\\
     131 &  3.7  &  4.2 &  3.6 &  3.3 &  2.8 &  3.5$\pm$ 0.5\\
     171 &  5.2  &  4.2 &  3.0 &  2.4 &  2.8 &  3.5$\pm$ 1.2\\
     193 &  4.9  &  3.8 &  2.9 &  2.9 &  3.0 &  3.5$\pm$ 0.9\\
     211 &  3.7  &  2.6 &  2.3 &  2.2 &  2.5 &  2.7$\pm$ 0.6\\
     304 &  3.9  &  3.1 &  2.9 &  2.6 &  2.3 &  2.9$\pm$ 0.6\\
     335 &  3.4  &  3.3 &  3.2 &  2.4 &  3.2 &  3.1$\pm$ 0.4\\
         &       &       &       &       &       &          \\
All  &  4.0$\pm$ 0.7 &  3.6$\pm$ 0.6 &  2.9$\pm$ 0.5 &  
	2.7$\pm$ 0.4 &  2.8$\pm$ 0.3 &  3.2$\pm$ 0.7 \\
\enddata
\end{deluxetable}

%__________________________TABLE 2________________________
\begin{deluxetable}{rrrrrrr}
\tabletypesize{\normalsize}
\tabletypesize{\footnotesize}
\tablecaption{Statistics of powerlaw slopes $\alpha_A$ 
of flare areas A[Mm$^2$] of
solar flares, tabulated for 7 different AIA wavelengths and 5 different
flux thresholds. The theoretical prediction
of the FD-SOC model is $\alpha_A=2 $.}
\tablewidth{0pt}
\tablehead{
\colhead{Wavelength}&
\colhead{Threshold}&
\colhead{Threshold}&
\colhead{Threshold}&
\colhead{Threshold}&
\colhead{Threshold}&
\colhead{All}\\
\colhead{[$\ang$]}&
\colhead{1\%}&
\colhead{2\%}&
\colhead{5\%}&
\colhead{10\%}&
\colhead{20\%}&
\colhead{}}
\startdata
      94 &  2.0 &  2.2 &  1.8 &  2.0 &  2.1 &  2.0$\pm$ 0.1\\
     131 &  2.4 &  2.2 &  2.2 &  2.2 &  1.9 &  2.2$\pm$ 0.2\\
     171 &  2.8 &  2.6 &  1.9 &  1.6 &  1.7 &  2.1$\pm$ 0.5\\
     193 &  2.6 &  2.1 &  1.8 &  1.8 &  1.7 &  2.0$\pm$ 0.3\\
     211 &  2.4 &  2.2 &  1.8 &  1.9 &  1.6 &  2.0$\pm$ 0.4\\
     304 &  2.2 &  2.3 &  2.1 &  2.1 &  1.9 &  2.1$\pm$ 0.2\\
     335 &  2.1 &  2.2 &  1.8 &  2.0 &  1.7 &  1.9$\pm$ 0.2\\
         &       &       &       &       &       &         \\
All  &  2.4$\pm$ 0.3 &  2.2$\pm$ 0.2 &  1.9$\pm$ 0.2 &  
	2.0$\pm$ 0.2 &  1.8$\pm$ 0.2 &  2.1$\pm$ 0.3 \\
\enddata
\end{deluxetable}

%__________________________TABLE 3________________________
\begin{deluxetable}{rrrrrrr}
\tabletypesize{\normalsize}
\tabletypesize{\footnotesize}
\tablecaption{Statistics of powerlaw slopes $\alpha_V$ 
of flare volumes V[Mm$^2$] of
solar flares, tabulated for 7 different AIA wavelengths and 5 different
flux thresholds. The theoretical prediction
of the FD-SOC model is $\alpha_V=5/3\approx 1.67$.}
\tablewidth{0pt}
\tablehead{
\colhead{Wavelength}&
\colhead{Threshold}&
\colhead{Threshold}&
\colhead{Threshold}&
\colhead{Threshold}&
\colhead{Threshold}&
\colhead{All}\\
\colhead{[$\ang$]}&
\colhead{1\%}&
\colhead{2\%}&
\colhead{5\%}&
\colhead{10\%}&
\colhead{20\%}&
\colhead{}}
\startdata
      94 &  1.6 &  1.4 &  1.5 &  1.6 &  1.6 &  1.5$\pm$ 0.1\\
     131 &  1.9 &  1.9 &  1.7 &  1.7 &  1.4 &  1.7$\pm$ 0.2\\
     171 &  2.0 &  1.8 &  1.5 &  1.4 &  1.5 &  1.7$\pm$ 0.2\\
     193 &  2.0 &  1.9 &  1.5 &  1.5 &  1.6 &  1.7$\pm$ 0.2\\
     211 &  1.8 &  1.7 &  1.6 &  1.6 &  1.4 &  1.6$\pm$ 0.2\\
     304 &  1.8 &  1.8 &  1.7 &  1.8 &  1.4 &  1.7$\pm$ 0.1\\
     335 &  1.6 &  1.6 &  1.6 &  1.5 &  1.7 &  1.6$\pm$ 0.1\\
         &       &       &       &       &       &         \\
All  &  1.8$\pm$ 0.2 &  1.7$\pm$ 0.2 &  1.6$\pm$ 0.1 &  
	1.6$\pm$ 0.1 &  1.5$\pm$ 0.1 &  1.6$\pm$ 0.2 \\
\enddata
\end{deluxetable}

%__________________________TABLE 4________________________
\begin{deluxetable}{rrrrrrr}
\tabletypesize{\normalsize}
\tabletypesize{\footnotesize}
\tablecaption{The diffusion coefficient $\kappa$ [Mm s$^{-\beta/2}$]
of 155 solar flares is tabulated for 7 different AIA wavelengths and 
5 different flux thresholds.}
\tablewidth{0pt}
\tablehead{
\colhead{Wavelength}&
\colhead{Threshold}&
\colhead{Threshold}&
\colhead{Threshold}&
\colhead{Threshold}&
\colhead{Threshold}&
\colhead{All}\\
\colhead{[$\ang$]}&
\colhead{1\%}&
\colhead{2\%}&
\colhead{5\%}&
\colhead{10\%}&
\colhead{20\%}&
\colhead{}}
\startdata
      94 &   19 &   21 &   22 &   23 &   23 &   21$\pm$   2\\
     131 &   23 &   25 &   28 &   30 &   31 &   27$\pm$   3\\
     171 &   46 &   41 &   36 &   35 &   35 &   39$\pm$   5\\
     193 &   43 &   40 &   39 &   39 &   41 &   40$\pm$   2\\
     211 &   33 &   30 &   30 &   31 &   32 &   31$\pm$   1\\
     304 &   29 &   27 &   28 &   28 &   28 &   28$\pm$   1\\
     335 &   19 &   19 &   20 &   21 &   20 &   20$\pm$   1\\
         &       &       &       &       &       &  \\
All &  30$\pm$ 11 &  29$\pm$  9 &  29$\pm$  7 &  
       30$\pm$  6 &  30$\pm$  7 &  30$\pm$  8 \\
\enddata
\end{deluxetable}

%__________________________TABLE 5________________________
\begin{deluxetable}{rrrrrrr}
\tabletypesize{\normalsize}
\tabletypesize{\footnotesize}
\tablecaption{The diffusion exponent $\beta$ of 155 solar flares
is tabulated for 7 different AIA wavelengths and 5 different
flux thresholds.  The theoretical prediction
of the FD-SOC model is $\beta=1$ for classical diffusion.}
\tablewidth{0pt}
\tablehead{
\colhead{Wavelength}&
\colhead{Threshold}&
\colhead{Threshold}&
\colhead{Threshold}&
\colhead{Threshold}&
\colhead{Threshold}&
\colhead{All}\\
\colhead{[$\ang$]}&
\colhead{1\%}&
\colhead{2\%}&
\colhead{5\%}&
\colhead{10\%}&
\colhead{20\%}&
\colhead{}}
\startdata
      94 &  0.44 &  0.45 &  0.45 &  0.42 &  0.39 &  0.43$\pm$ 0.03\\
     131 &  0.26 &  0.27 &  0.28 &  0.26 &  0.16 &  0.25$\pm$ 0.05\\
     171 &  0.48 &  0.47 &  0.43 &  0.39 &  0.37 &  0.43$\pm$ 0.05\\
     193 &  0.36 &  0.24 &  0.18 &  0.16 &  0.19 &  0.23$\pm$ 0.08\\
     211 &  0.47 &  0.06 &  0.42 &  0.43 &  0.28 &  0.33$\pm$ 0.17\\
     304 &  0.36 &  0.34 &  0.32 &  0.30 &  0.03 &  0.27$\pm$ 0.14\\
     335 &  0.56 &  0.50 &  0.49 &  0.52 &  0.44 &  0.50$\pm$ 0.04\\
         &       &       &       &       &       &                \\
All &  0.42$\pm$ 0.10 &  0.33$\pm$ 0.16 &  0.37$\pm$ 0.11 &  
       0.35$\pm$ 0.12 &  0.27$\pm$ 0.15 &  0.35$\pm$ 0.13 \\
\enddata
\end{deluxetable}

%__________________________TABLE 6________________________
\begin{deluxetable}{rrrrrrr}
\tabletypesize{\normalsize}
\tabletypesize{\footnotesize}
\tablecaption{Statistics of fractal dimensions $D_2$  
of solar flare areas 
tabulated for 7 different AIA wavelengths and 5 different
flux thresholds.  The theoretical prediction
of the FD-SOC model is $D_2=3/2$.}
\tablewidth{0pt}
\tablehead{
\colhead{Wavelength}&
\colhead{Threshold}&
\colhead{Threshold}&
\colhead{Threshold}&
\colhead{Threshold}&
\colhead{Threshold}&
\colhead{All}\\
\colhead{[$\ang$]}&
\colhead{1\%}&
\colhead{2\%}&
\colhead{5\%}&
\colhead{10\%}&
\colhead{20\%}&
\colhead{}}
\startdata
      94 &  1.7 &  1.7 &  1.6 &  1.6 &  1.5 &  1.6$\pm$ 0.1\\
     131 &  1.7 &  1.7 &  1.6 &  1.6 &  1.5 &  1.6$\pm$ 0.1\\
     171 &  1.7 &  1.6 &  1.5 &  1.4 &  1.3 &  1.5$\pm$ 0.1\\
     193 &  1.7 &  1.7 &  1.6 &  1.6 &  1.5 &  1.6$\pm$ 0.1\\
     211 &  1.7 &  1.6 &  1.6 &  1.5 &  1.4 &  1.6$\pm$ 0.1\\
     304 &  1.6 &  1.6 &  1.5 &  1.4 &  1.2 &  1.5$\pm$ 0.2\\
     335 &  1.6 &  1.6 &  1.5 &  1.4 &  1.3 &  1.5$\pm$ 0.1\\
         &       &       &       &       &       &         \\
All  &  1.7$\pm$ 0.1 &  1.6$\pm$ 0.1 &  1.6$\pm$ 0.1 &  
	1.5$\pm$ 0.1 &  1.4$\pm$ 0.1 &  1.6$\pm$ 0.1 \\
\enddata
\end{deluxetable}

%__________________________TABLE 7________________________
\begin{deluxetable}{lrrlllll}
\tabletypesize{\normalsize}
\tabletypesize{\footnotesize}
\tablecaption{Size distributions of geometric and temporal
parameters in solar flares.}
\tablewidth{0pt}
\tablehead{
\colhead{Instrument}&
\colhead{Wavelength}&
\colhead{Number}&
\colhead{Length}&
\colhead{Area}&
\colhead{Volume}&
\colhead{Duration}&
\colhead{Reference}\\
\colhead{}&
\colhead{or energy}&
\colhead{of events}&
\colhead{exponent}&
\colhead{exponent}&
\colhead{exponent}&
\colhead{exponent}&
\colhead{}\\
\colhead{}&
\colhead{$\lambda$, $\epsilon$}&
\colhead{$N$}&
\colhead{$\alpha_L$}&
\colhead{$\alpha_A$}&
\colhead{$\alpha_V$}&
\colhead{$\alpha_T$}&
\colhead{}}
\startdata
SMM/HXRBS    &$>$25 keV  & 7045 &               &               &               & 2.17$\pm$0.05 & Crosby et al.~(1993) \\ %1980-1982
SMM/HXRBS    &$>$25 keV  & 1008 &               &               &               & 1.95$\pm$0.09 & Crosby et al.~(1993) \\ %1983-1984
SMM/HXRBS    &$>$25 keV  &  545 &               &               &               & 2.22$\pm$0.13 & Crosby et al.~(1993) \\ %1985-1987
SMM/HXRBS    &$>$25 keV  & 3874 &               &               &               & 1.99$\pm$0.06 & Crosby et al.~(1993) \\ %1988-1989
ISEE-3       &$>$25 keV  & 4356 &               &               &               & 1.88          & Lu et al.~(1993)     \\
ISEE-3       &$>$25 keV  & 4356 &               &               &               & 2.73          & Lee et al.~(1993)    \\
ISEE-3       &$>$25 keV  & 4356 &               &               &               & 2.40$\pm$0.04 & Bromund et al.~(1995)\\
GRANAT/WATCH &$>$10 keV  & 1551 &               &               &               & 1.09-2.28\tablenotemark{a} & Crosby et al.~(1998) \\ 
SOHO/EIT     & 304 \ang  &13,067 &               & 2.7           &               & 3.1           & Berghmans et al.~(1998)\\
SOHO/EIT     & 195 \ang  &13,607 &               & 2.0           &               & 2.1           & Berghmans et al.~(1998)\\
TRACE        &171-195 \ang &   281 & 2.10$\pm$0.11 & 2.56$\pm0.23$ & 1.94$\pm0.09$ &               & Aschwanden et al. (2000)\\
GRANAT/WATCH &$>$10 keV  & 1518 \ang &               &               &               & 1.15-2.25\tablenotemark{a} & Georgoulis et al.~(2001)\\  
TRACE/C      &171-195 \ang &     & 3.24$\pm$0.16 & 2.43$\pm$0.10 & 2.08$\pm$0.07 &               & Aschwanden and Parnell (2002)\\ %Parnell-code
TRACE/A      & 171 \ang  &   436 & 2.87$\pm$0.24 & 2.45$\pm$0.09 & 1.65$\pm$0.09 &               & Aschwanden and Parnell (2002)\\
TRACE/B      & 171 \ang  &   436 & 2.77$\pm$0.17 & 2.34$\pm$0.10 & 1.75$\pm$0.13 &               & Aschwanden and Parnell (2002)\\
TRACE/A      & 195 \ang  &   380 & 2.59$\pm$0.19 & 2.16$\pm$0.18 & 1.69$\pm$0.05 &               & Aschwanden and Parnell (2002)\\
TRACE/B      & 195 \ang  &   380 & 2.56$\pm$0.17 & 2.24$\pm$0.04 & 1.63$\pm$0.04 &               & Aschwanden and Parnell (2002)\\
Yohkoh/SXT   & AlMg      &   103 & 2.34$\pm$0.27 & 1.86$\pm$0.13 & 1.44$\pm$0.07 &               & Aschwanden and Parnell (2002)\\
TRACE+SXT &171,195,AlMg  &   919 & 2.41$\pm$0.09 & 1.94$\pm$0.03 & 1.55$\pm$0.03 &               & Aschwanden and Parnell (2002)\\
GOES         & 1-8 \ang  & 49,409&               &               &               & 2.93$\pm$0.12\tablenotemark{c} & Veronig et al.~(2002)\\ %(no backgr subtr)
SONO/EIT     & 195 \ang  &200,000&               &               &               & 1.4-2.0\tablenotemark{b}   & McIntosh and Gurman (2005)\\ 
RHESSI       &$12-25$keV &2649 &                 &               &               & 0.9-3.56\tablenotemark{c}  & Su et al.~(2006)     \\
GOES         & 1-8 \ang  &  1365 &               &               &               & 2.87$\pm$0.09\tablenotemark{c} & Yashiro et al.~(2006) \\ %(no backgr subtr)
GOES         & 1-8 \ang  &       &               &               &               & 2.49$\pm$0.11\tablenotemark{c} & Yashiro et al.~(2006) \\ %(with CME)
GOES         & 1-8 \ang  &       &               &               &               & 3.22$\pm$0.15\tablenotemark{c} & Yashiro et al.~(2006) \\ %(w/o CME)
RHESSI       & 6-12 keV  & 25705 &               &               &               & 2.2$\pm$0.18  & Christe et al.~(2008)\\
TRACE        & 1550 \ang &       &               &               &               & 2.3           & Nishizuka et al.~(2009)\\
SMM/HXRBS    & $>$25 keV & 11,549&               &               &               & 2.05          & Aschwanden (2011)    \\
CGRO/BATSE   & $>$25 keV &   4109&               &               &               & 2.20          & Aschwanden (2011)    \\
RHESSI       & $>$25 keV & 11,595&               &               &               & 2.00          & Aschwanden (2011)    \\
GOES	     & 1-8 \ang  &255,474&               &               &               & 3.20$\pm0.04$\tablenotemark{d} & Aschwanden and Freeland (2012)\\ %solar max
GOES	     & 1-8 \ang  & 67,296&               &               &               & 2.75$\pm0.02$\tablenotemark{e} & Aschwanden and Freeland (2012)\\ %intermediate
GOES	     & 1-8 \ang  & 18,676&               &               &               & 2.26$\pm0.01$\tablenotemark{f} & Aschwanden and Freeland (2012)\\ %solar minimum
GOES         & 1-8 \ang  &   155 &               &               &               & 1.92          & Aschwanden (2012b)   \\
AIA/SDO      & 335 \ang  &   155 & 1.96          &               &               & 2.17          & Aschwanden (2012b)   \\
GOES         & 1-8 \ang  &   155 &               &               &               & 2.10$\pm$0.18 & This work            \\
AIA/SDO      &  94 \ang  &   155 & 3.1$\pm$0.6 & 2.0$\pm$0.1 & 1.5$\pm$0.1 &               & This work            \\
AIA/SDO      & 131 \ang  &   155 & 3.5$\pm$0.5 & 2.2$\pm$0.2 & 1.7$\pm$0.2 &               & This work            \\
AIA/SDO      & 171 \ang  &   155 & 3.5$\pm$1.2 & 2.1$\pm$0.5 & 1.7$\pm$0.2 &               & This work            \\
AIA/SDO      & 193 \ang  &   155 & 3.5$\pm$0.9 & 2.0$\pm$0.3 & 1.7$\pm$0.2 &               & This work            \\
AIA/SDO      & 211 \ang  &   155 & 2.7$\pm$0.6 & 2.1$\pm$0.3 & 1.6$\pm$0.2 &               & This work            \\
AIA/SDO      & 304 \ang  &   155 & 2.9$\pm$0.6 & 2.1$\pm$0.2 & 1.7$\pm$0.1 &               & This work            \\
AIA/SDO      & 335 \ang  &   155 & 3.1$\pm$0.4 & 1.9$\pm$0.2 & 1.6$\pm$0.1 &               & This work            \\
AIA/SDO      & 94-335 \ang & 155 & 3.2$\pm$0.7 & 2.1$\pm$0.3 & 1.6$\pm$0.2 &               & This work            \\
Theory	     &        &       & {\bf 3.00}    & {\bf 2.00}    & {\bf 1.67}    & {\bf 2.00}    &                      \\
\enddata
\tablenotetext{a}{Double-powerlaw fit.}
\tablenotetext{b}{Powerlaw fit with exponential fall-off.}
\tablenotetext{c}{No preflare background flux subtraction.}
\tablenotetext{d}{During solar cycle minima years.}
\tablenotetext{e}{During intermediate solar cycle years.}
\tablenotetext{f}{During solar cycle maxima years.}
\end{deluxetable}

\clearpage
%%%%%%%%%%%%%%%%%%%%%%%%%%% FIGURE %%%%%%%%%%%%%%%%%%%%%%%%%%%%%%%%% 

\begin{figure}
\plotone{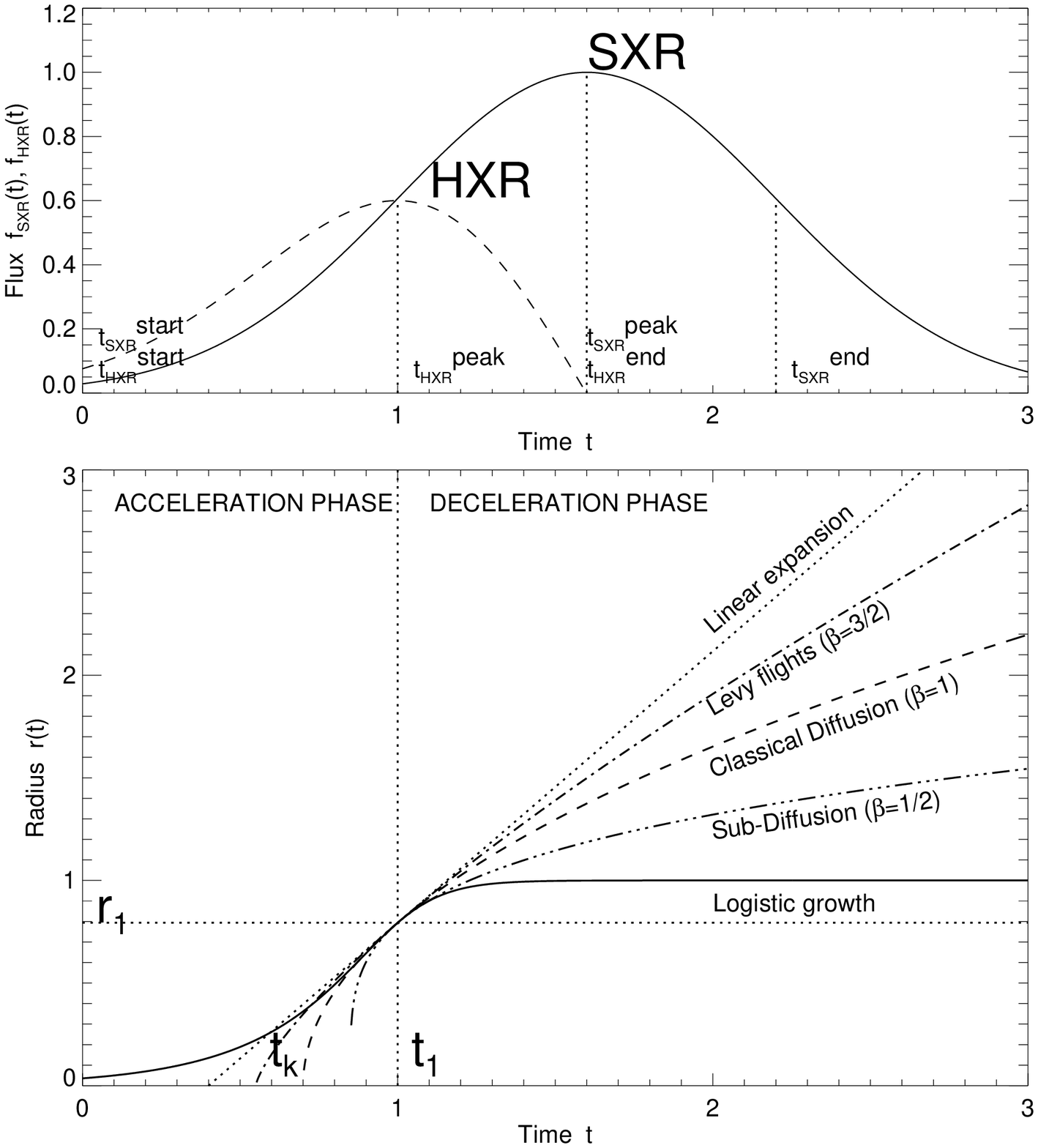}
\caption{{\sl Top:} Time evolution of the soft X-ray flux $f_{SXR}(t)$
and hard X-ray flux $f_{HXR}(t)\approx df_{SXR}/dt$ during a flare,
following the Neupert effect. The rise time of the SXR flux,
$t_{SXR}^{peak}-t_{SXR}^{start}$ coincides with the duration of
the hard X-ray emission $T=t_{HXR}^{end}-t_{HXR}^{start}=
t_{SXR}^{peak}-t_{SXR}^{start}$.
{\sl Bottom:} Comparison of spatio-temporal evolution models:
Logistic growth with parameters $t_L=1.0, r_\infty=1.0, \tau_G=0.1$,
sub-diffusion ($\beta=1/2$), classical diffusion ($\beta=1$),
L\'evy flights or hyper-diffusion ($\beta=3/2$), and linear expansion
($r \propto t$).  All three curves intersect at $t=t_L$
and have the same speed $v=(dr/dt)$ at the intersection point at
time $t=t_L$ (adapted from Aschwanden 2012b).}
\end{figure}

\begin{figure}
\plotone{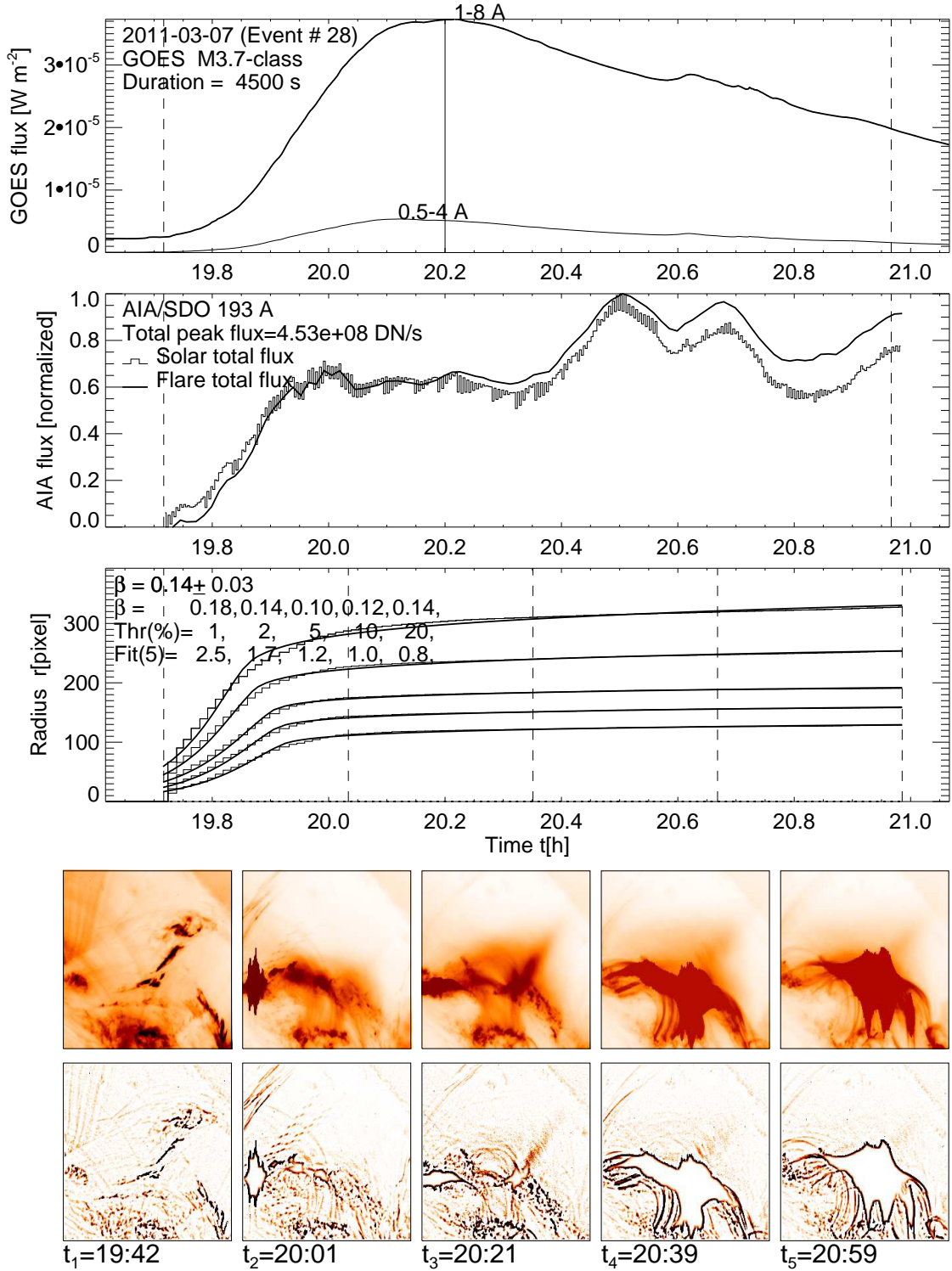}
\caption{Flare event \#28 was observed on 2011 March 7, 19:43-20:58 UT,
with AIA/SDO 193 \ang . The panels show the GOES time profiles (top panel),
the total flux (second panel), the spatio-temporal evolution of the
radius $r_{193,th}(t)=\sqrt{A(t)/\pi}$ of the time-integrated flare 
area $A(t)$ for 5 different thresholds, i.e., 1\%, 2\%, 5\%, 10\%, 20\% of
the peak flux (third panel: histogrammed), fitted with the anomalous
diffusion model (third panel: solid curves), and five snapshots of the
193 \ang\ flux (fourth row of panels), and highpass-filtered flux
(bottom row of panels).}
\end{figure}

\begin{figure}
\plotone{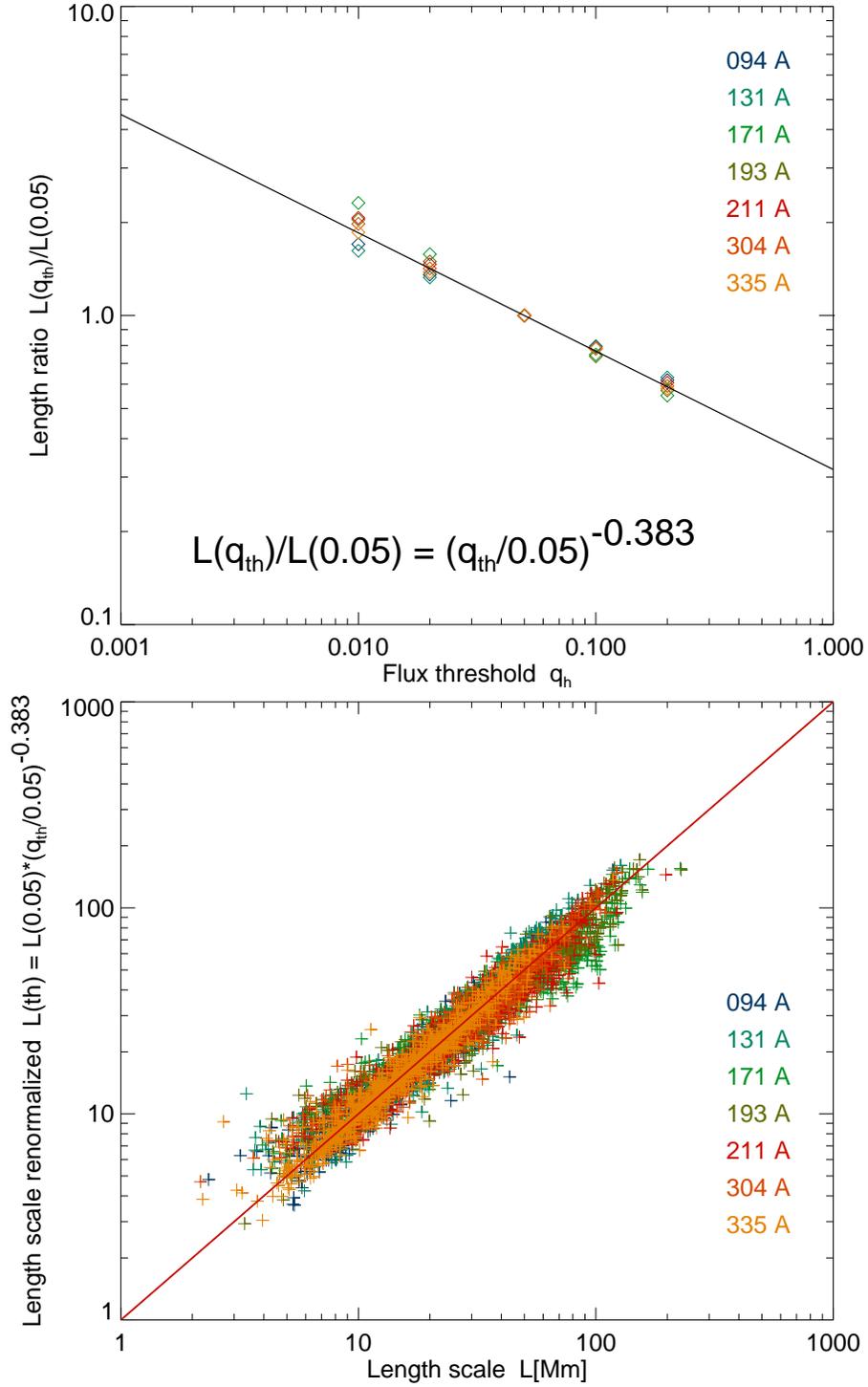}
\caption{The normalization of the length scale $L$ for different
thresholds $q_{th}=0.01, 0.02, 0.05, 0.1, 0.2$ is shown for 7 different
wavelength filters (different colors). The mean ratios of the 
measured length scales $L(q_{th})/L(0.05)$ as a function of the threshold 
$q_th$ is shown in the top panel, and the renormalized length scales
$L_\lambda(q_{th})=L_\lambda(0.05) \times (q_{th}/0.05)^{-0.383}$
are shown in the bottom panel for all 155 flares and 7 wavelengths 
(different colors).}
\end{figure}

\begin{figure}
\plotone{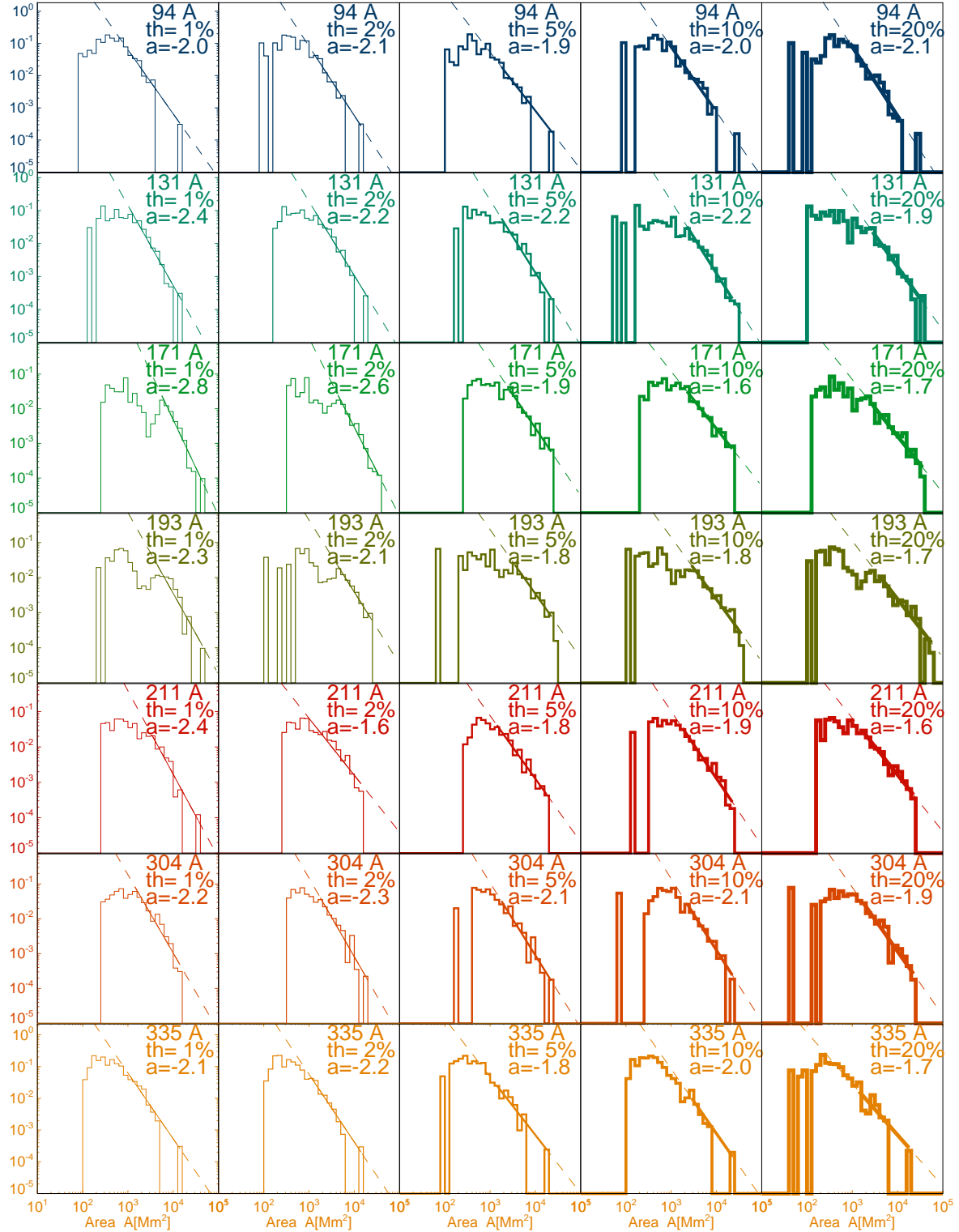}
\caption{Log-log histograms $N(A)$ of measured flare areas $A$ are shown 
for the dataset of 155 analyzed events, for 7 different wavelengths (rows;
marked with identical color) and for 5 different flux thresholds (columns; 
marked with line thickness increasing as a function of the flux threshold.
Powerlaw fits are obtained in the logarithmic area range of 
$0.05 A_{max} \le A \le A_{max}$, with the powerlaw slopes indicated in 
each panel and listed in Table 2.}
\end{figure}

\begin{figure}
\plotone{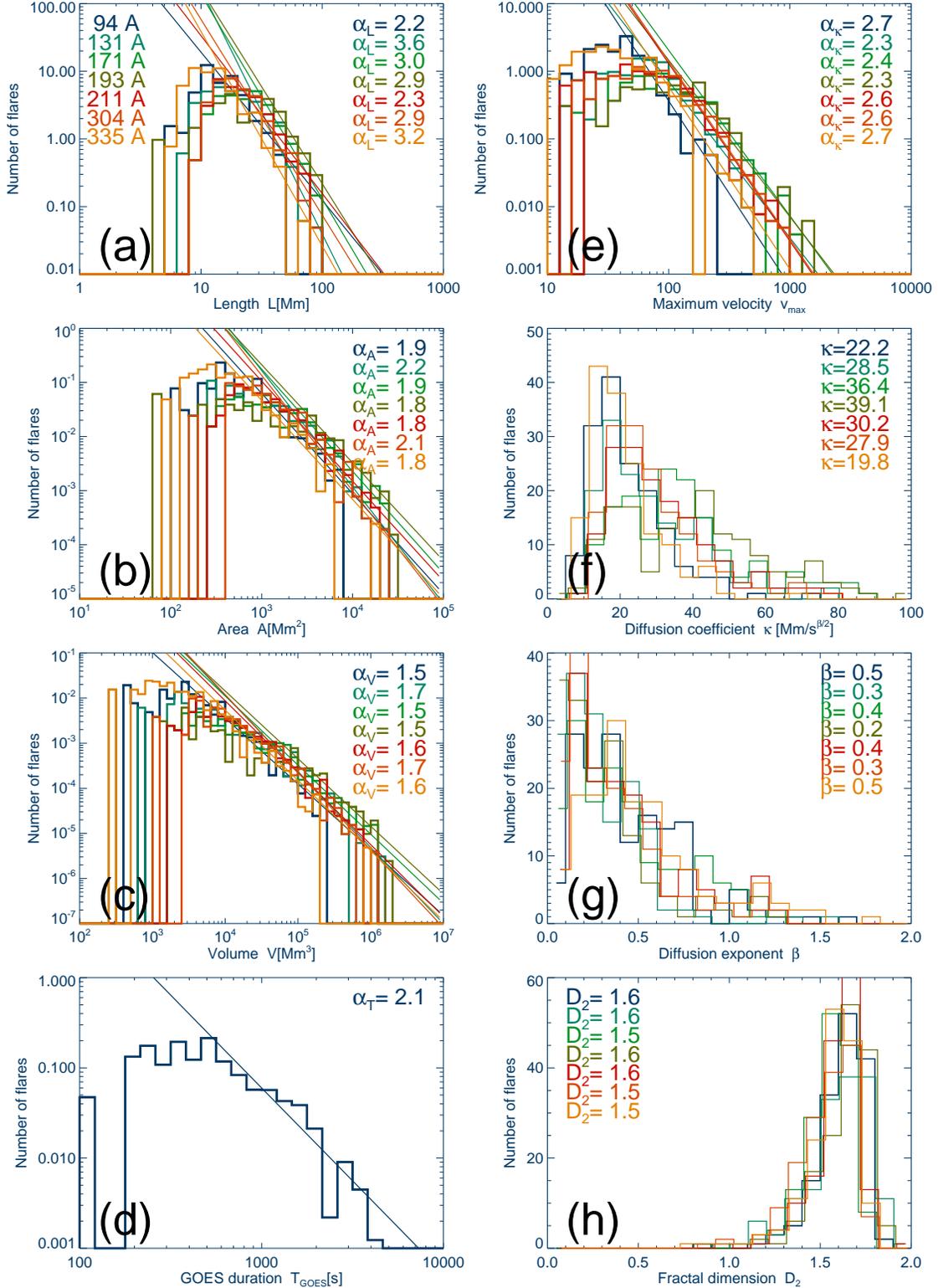}
\caption{Log-log histograms of length scales $l$, areas $A$, volumes $V$,
GOES flare durations $T$, diffusion coefficients $\kappa$, diffusion
exponents $\beta$, maximum velocity $v_{max}$, and 2-D fractal dimensions
 $D_2$ sampled from the 155 analyzed flares, in each of the 7 wavelengths
(94, 131, 171, 193, 211, 304, 335 \ang\ , marked with different colors),
for a threshold of $q_{th}=5\%$. The fitted powerlaw slopes are indicated
for each wavelength separately in each panel.}
\end{figure}

\begin{figure}
\plotone{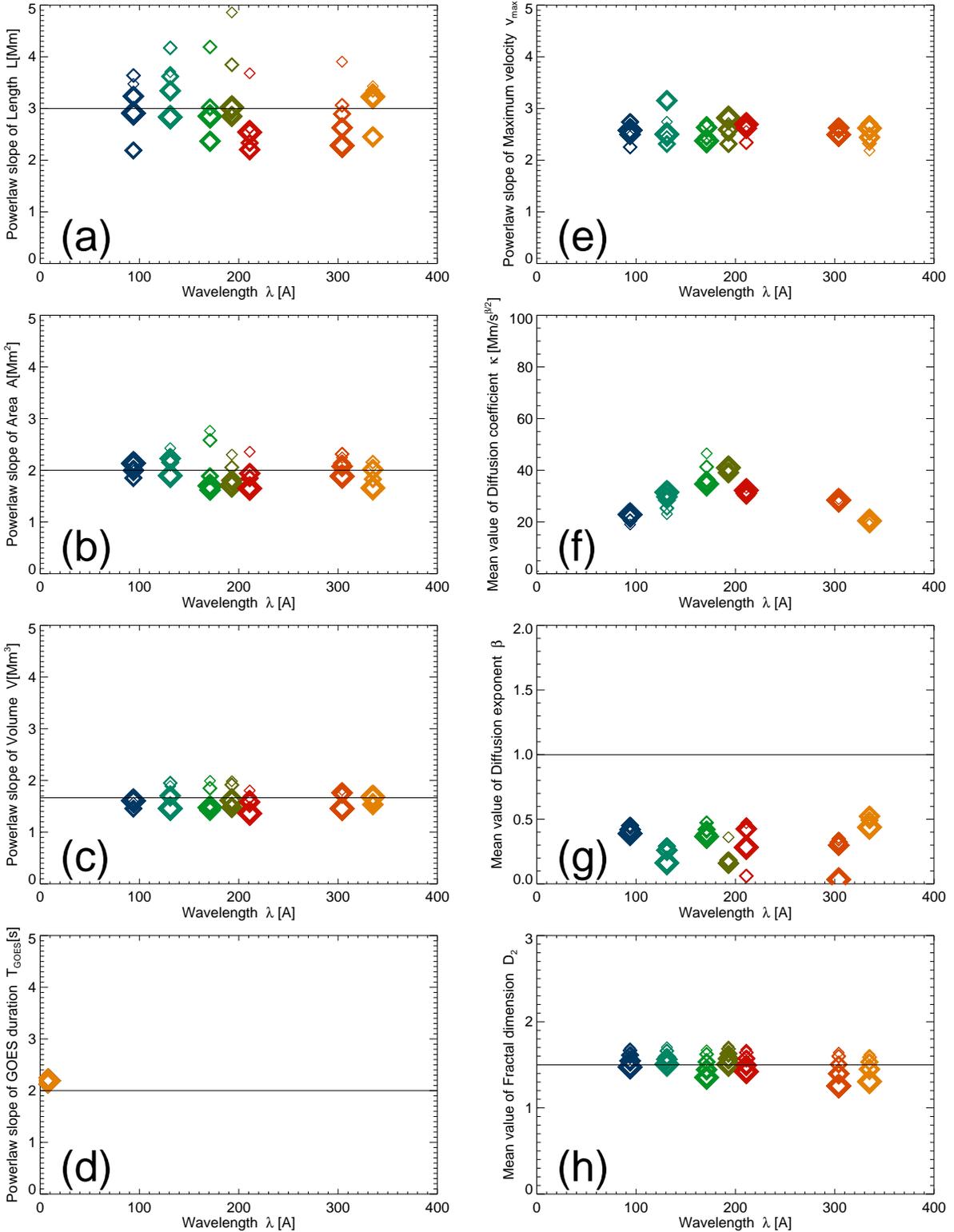}
\caption{Powerlaw indices $\alpha$ and fractal dimension $D_2$ as a 
function of the wavelength $\lambda$ (with different colors). 
The size of the diamond symbols indicate the t flux threshold levels.
The values predicted by the FD-SOC model are indicated with a 
horizontal line.}
\end{figure}

\begin{figure}
\plotone{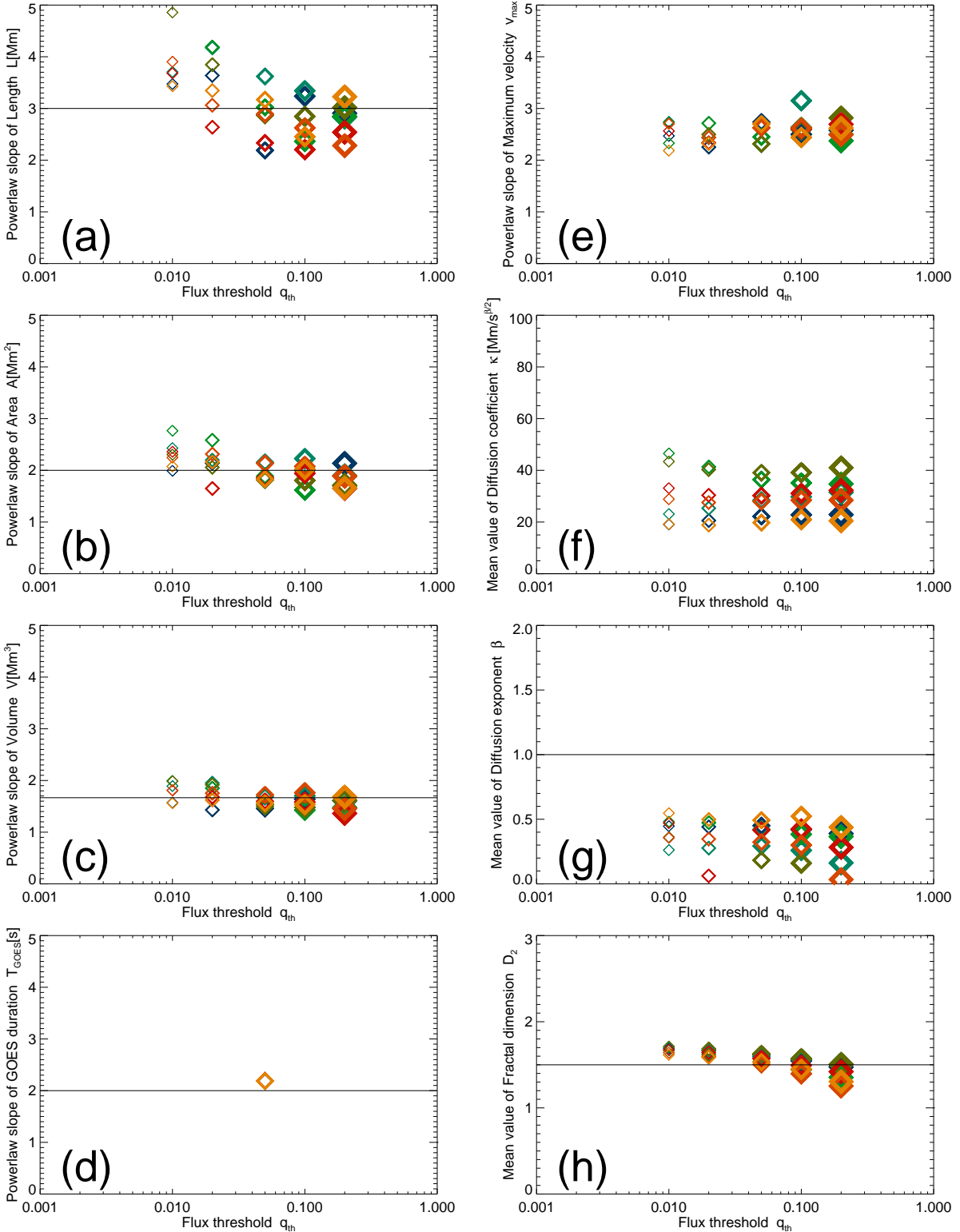}
\caption{Powerlaw indices $\alpha$ and fractal dimension $D_2$ as a 
function of the flux threshold $q_{th}$ (with different diamond sizes). 
The color indicates the wavelength. The values predicted by the FD-SOC 
model are indicated with a horizontal line.}
\end{figure}

\begin{figure}
\plotone{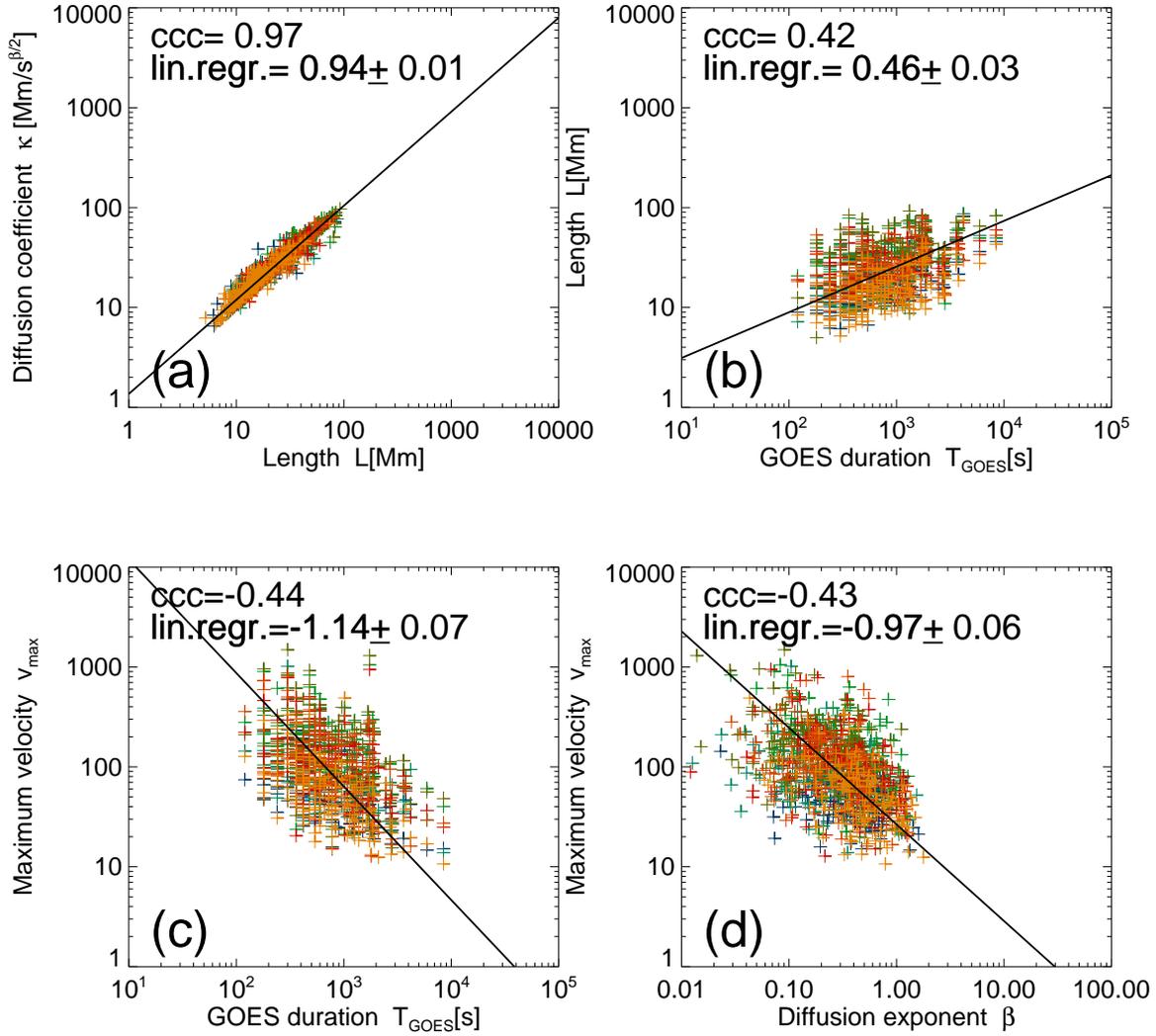}
\caption{Scatterplots between correlated parameters of the length scale $L$,
the (GOES) time duration $T$, the diffusion exponent $\beta$, and 
the maximum velocity $v_{max}$,
for 7 different wavelengths (in different colors) using a flux threshold 
of $q_{th}=5\%$. The Pearson cross-correlation coefficient ($ccc$) is
indicated and the linear regression slope with uncertainty.}
\end{figure}

\begin{figure}
\plotone{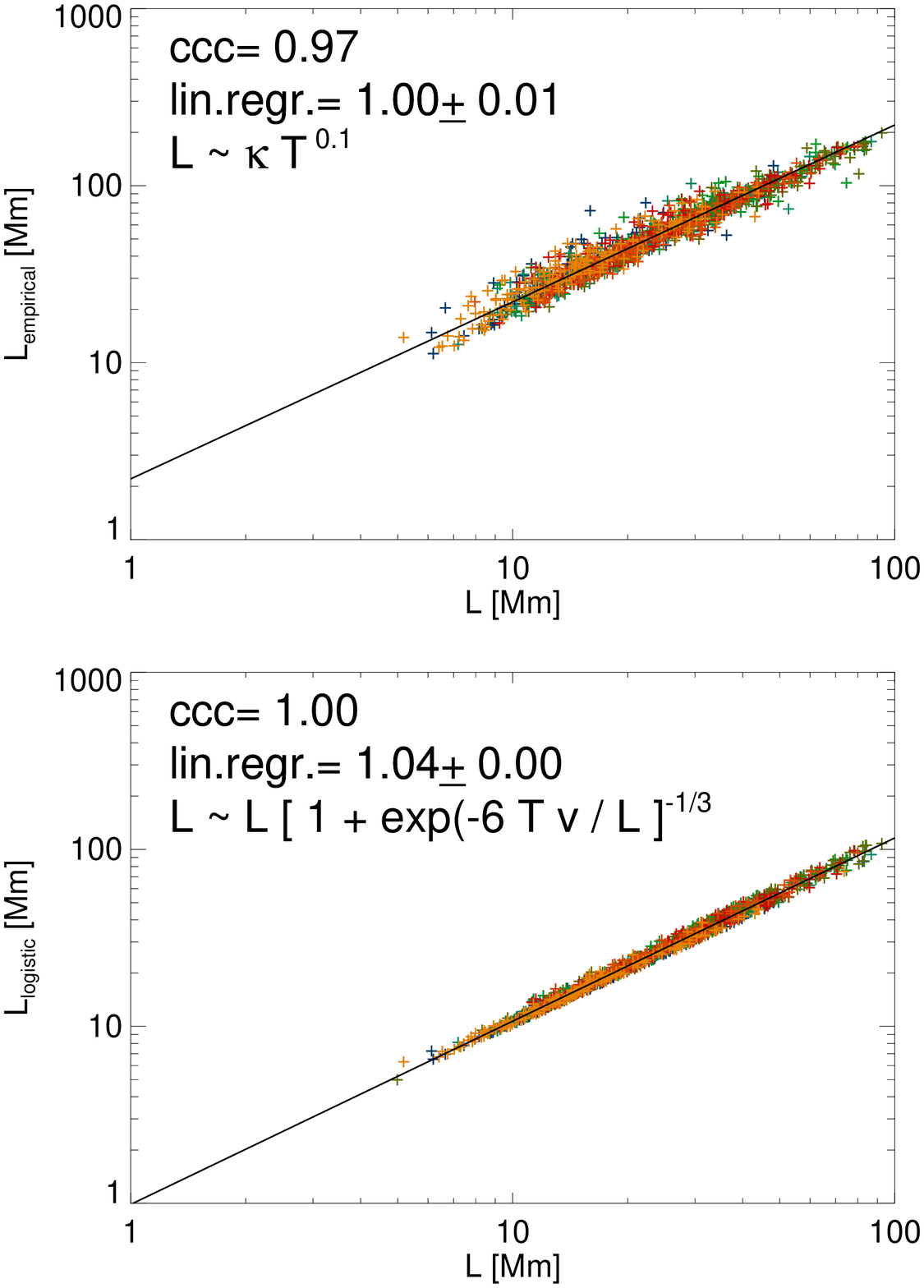}
\caption{{\sl Top:} The three-parameter correlation between the 
length scale $L$, the time scale $T$, and the diffusion coefficient 
$\kappa$ shows a best fit for the relationship $L \propto \kappa T^{0.1}$,
with a Pearson cross-correlation coefficient of $ccc = 0.97$.}
\end{figure}.  

\begin{figure}
\plotone{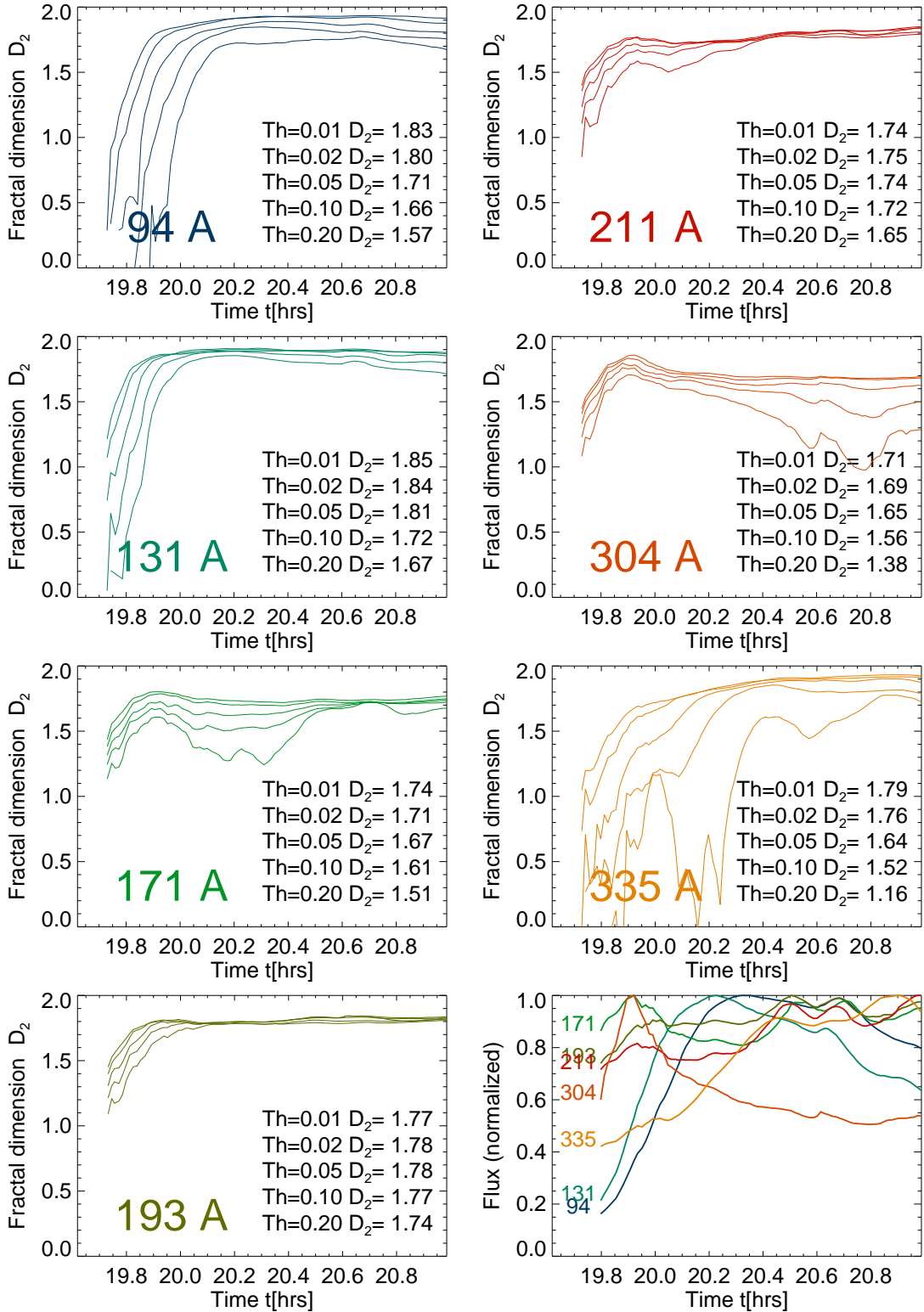}
\caption{2-D Fractal dimension $D_2(t)$ measured for events \#28,
2011 March 7 (see also Fig.~2), as a function of time during the flare,
shown for all 7 wavelengths (in different colors) and 5 different 
thresholds ($q_{th}=1\%, 2\%, 5\%, 10\%, 20\%$). The lowest threshold yields
the curve with the highest fractal dimension. For comparison, the
normalized total fluxes are also shown in the bottom right panel.}
\end{figure}

\begin{figure}
\plotone{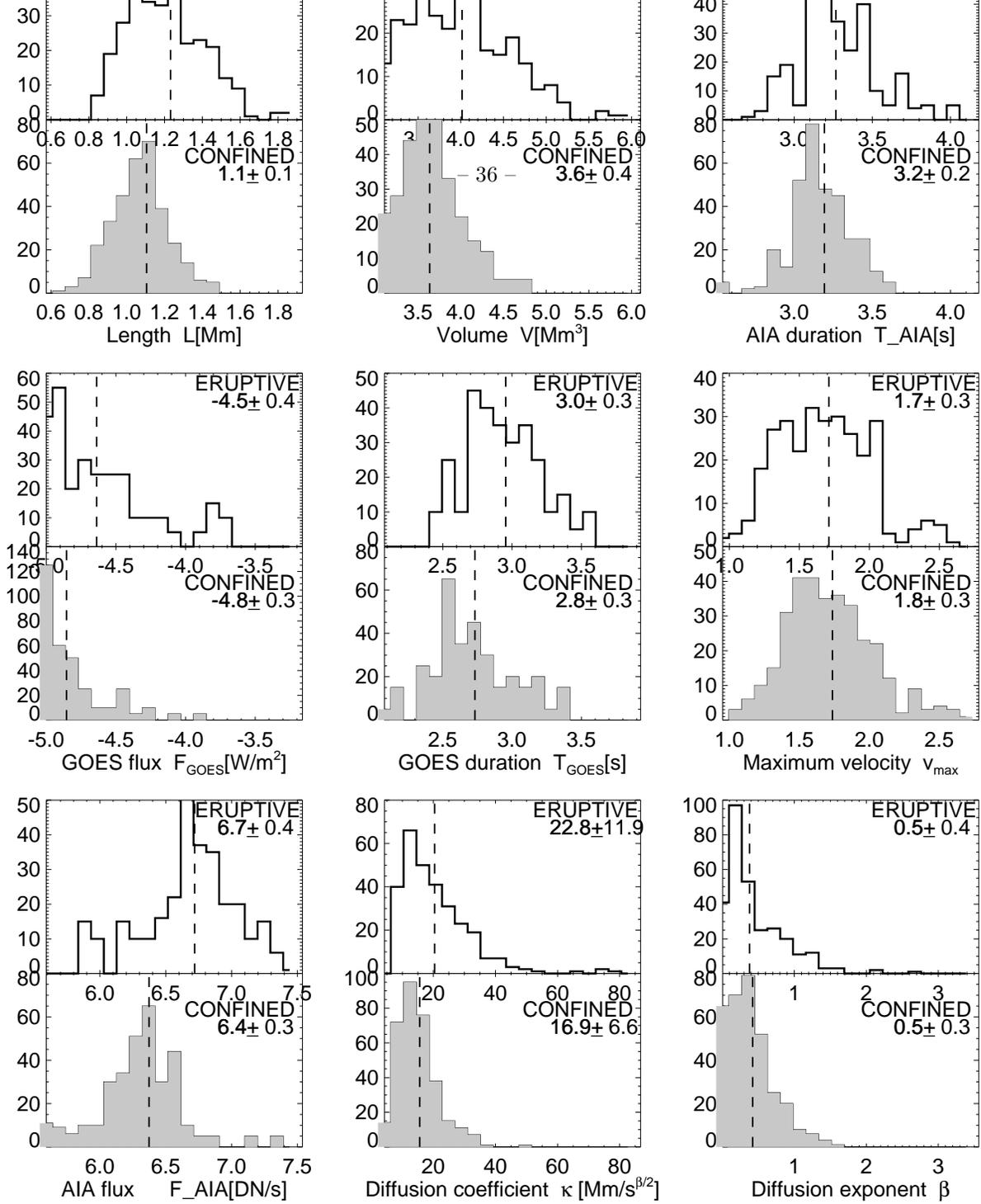}
\caption{Histograms of measured flare parameters of the
155 analyzed flares (at a wavelength of 335 \ang )
subdivided into the two subgroups of eruptive
flares (white histograms) and confined flares (grey histograms),
according to the classification of Zhang and Liu (2012). The mean values
and standard deviations are listed, and the mean value is marked 
with a vertical dashed bar.}
\end{figure}

\begin{figure}
\plotone{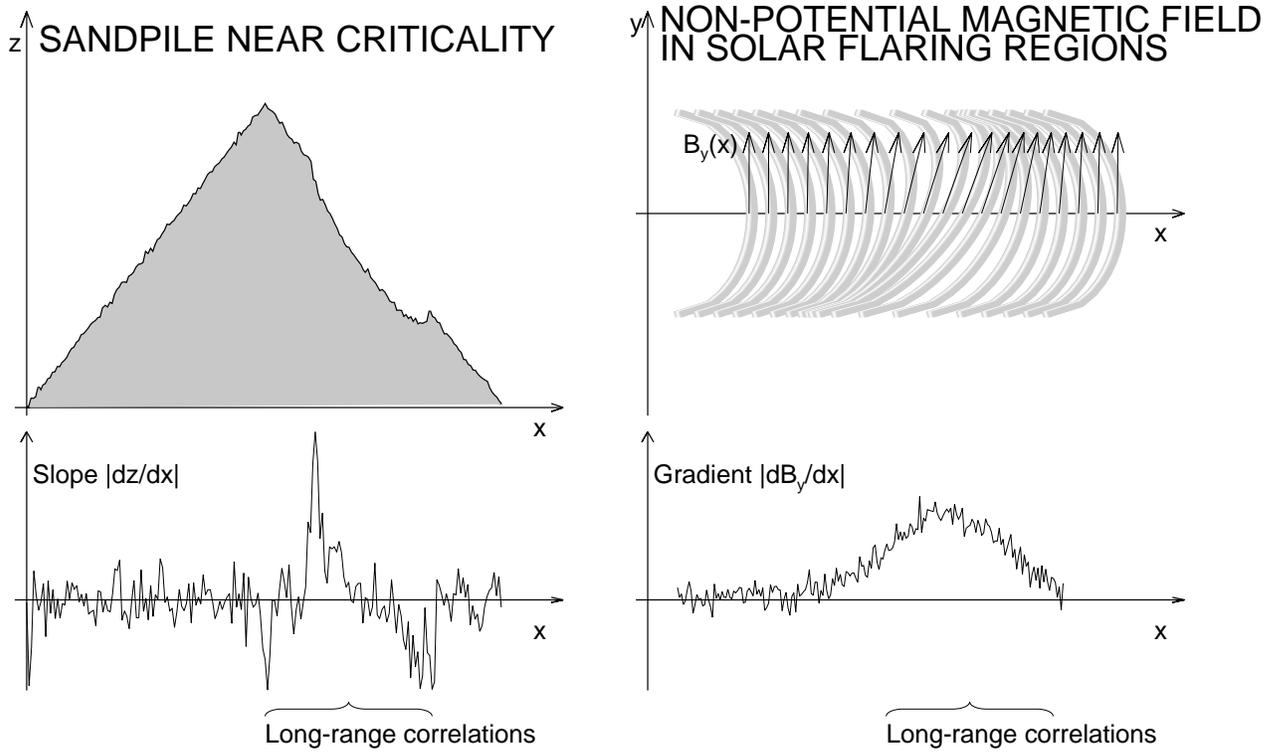}
\caption{{\sl Left:} A sandpile in a state in the vicinity of criticality
is shown with a height cross-section $z(x)$, with the gradient of the slope 
(or repose angle) $|dz/dx|$ (bottom), exhibiting short-range fluctuations
due to noise and long-range correlation lengths due to locally extended 
deviations
from the mean critical slope. {\sl Right:} The solar analogy of a flaring
region is visualized in terms of a loop arcade over a neutral line
in $x$-direction, consisting of loops with various shear angles that
are proportional to the gradient of the field direction $B_x/B_y$, 
showing also some locally extended (non-potential) deviations from
the potential field (bottom).}
\end{figure}

\end{document}